\documentstyle [11pt,psfig]{article}
\def\gs{\mathrel{\raise0.35ex\hbox{$\scriptstyle >$}\kern-0.6em 
\lower0.40ex\hbox{{$\scriptstyle \sim$}}}}
\def\ls{\mathrel{\raise0.35ex\hbox{$\scriptstyle <$}\kern-0.6em 
\lower0.40ex\hbox{{$\scriptstyle \sim$}}}}
\begin{document}
\pagenumbering{arabic}
\pagestyle{myheadings}
{\markright{Luminosity Distributions within rich clusters - III: ...}}
\title{Luminosity Distributions within Rich Clusters - III:
A comparative study of seven Abell/ACO clusters}
\author{Simon P. Driver$^{1}$, Warrick J. Couch$^{1}$ and 
Steven Phillipps$^{2}$\\
$^{1}$ School of Physics, University of New South Wales,\\ 
Sydney, NSW 2052, Australia,
$^{2}$ Astrophysics Group, University of Bristol, \\
Tyndall Avenue, BS8 1TL, UK}

\maketitle

\begin{abstract}
We recover the luminosity distributions over a wide range of absolute magnitude
($-24.5 < M_{R} < -16.5$) for a sample of seven rich southern galaxy clusters.
We find a large variation in the ratio of dwarf to giant galaxies, DGR: 
$0.8\le $ DGR $\le 3.1$. This variation is shown to be {\it inconsistent} with 
a ubiquitous cluster luminosity function. The DGR shows a smaller variation
from cluster to cluster 
in the inner regions ($r\ls 0.56$\,Mpc). Outside these regions 
we find the DGR to be strongly
anti--correlated with the mean local projected galaxy density with the 
DGR increasing towards lower densities. In addition the DGR in the outer
regions shows some correlation with Bautz--Morgan type. Radial analysis of 
the clusters indicate that the dwarf galaxies are less centrally clustered 
than the giants and form a significant halo around clusters. We conclude that  
measurements of the total cluster luminosity distribution based on the inner
core alone are likely to be severe underestimates of the dwarf component,
the integrated cluster luminosity and the contribution of
galaxy masses to the cluster's total mass. Further work is required to 
quantify this. The observational evidence that the unrelaxed, lower density
outer regions of clusters are dwarf--rich, adds credence to the recent 
evidence and conjecture that the field is a predominantly dwarf rich 
environment and that the dwarf galaxies are under--represented in measures of  
the local field luminosity function.

\end{abstract}

\section{Introduction}
Until recently the majority of studies of the luminosity distribution of 
galaxies within
rich clusters have concentrated on the brightest galaxies (e.g. Oemler 1974; 
Dressler 1978) and, in particular, the determination of any correlation 
between the brightest cluster member ($M_{1}$) and the turn-over point 
($M_{*}$) of the best fit Schechter function. 
However, despite extensive surveys, the identification of a clear trend in 
$M_{1}-M_{*}$ with any cluster property remains contentious (see
for example: Tr\`{e}vese, Crimele \& Appodia 1996; or Lumsden 
et al. 1997 and references therein for two differing views).
What does appear robust, however, is the lack of any observable change
in the mean cluster $M_{*}$ with redshift (for $z < 0.5$), indicating little 
luminosity evolution ($\Delta m < \pm 0.2$) of the brightest cluster members
in recent times (e.g. Dressler et al. 1994; Smail et al. 1997; Driver et al. 
1997; Schade, Barrientos \& L\^{o}pez-Cruz 1997; Barger et al. 1998). Is 
global environment irrelevant then to the 
luminosity function of galaxies and its evolution (at least over the 
observable range covered)? 
The presence of the Butcher-Oemler effect (Butcher \& Oemler 1978) observed
for most clusters at $z\gs 0.2$ suggests not. 
This effect has also now been linked, via high-resolution HST imaging, 
to a population of mostly late--type spirals (Couch et al. 1994; 
Oemler, Dressler \& Butcher 1997; Moore, Lake \&
Katz 1998; Couch et al. 1998). It seems then that the interesting changes
occurring in the galaxy populations within clusters are at sub-L$_{*}$ 
luminosities and therefore 
unlikely to be well traced by bright end luminosity function fitting.
Indeed most current models of cluster development (e.g. Charlot \& 
Silk 1994; Kauffmann, Nusser \& Steinmetz 1997) predict the rapid
formation of the cD and brightest galaxies followed by the slower evolution
of the later-types through processes such as galaxy harassment (Moore {\it et 
al.} 1996).

This state of play appears intriguingly analogous to that found for
field galaxies over a similar redshift range ($0.0 < z < 0.5$).
Three entirely independent methods of studying the evolution of field galaxies
indicate minimal change in the luminous population, 
these methods are: morphological galaxy counts (Driver, Windhorst \&
Griffiths 1995; Driver et al. 1995; Glazebrook et al. 1995;
Abraham et al. 1996; Odewahn et al. 1996); tracing the field
luminosity function via redshift surveys (Lilly et al. 1995; Colless
1995; Ellis et al. 1996) and the study of randomly identified high 
redshift MgII absorbers (Steidel, Dickinson \& Perrson 1994). The conclusion of
these studies
is that it is the lower luminosity population (late-type spirals, irregulars
and various dwarfs) which have most recently undergone evolutionary processes
(Driver et al. 1996). 
Is the faint blue galaxy problem (Koo \& Kron 1992; Ellis 1997) and the 
Butcher-Oemler effect the manifestation of the same process in different 
environments?

Given the lack of obvious change of the most luminous galaxies in {\it any}
environment and the numerous indications of recent evolution in the 
sub-L$_{*}$ population in both cluster environments {\it and} the field, it 
seems prudent to concentrate more on the properties of the lower luminosity 
galaxies within clusters and their likely end products.
For both the field and cluster environments the prediction is that 
these evolving late-type systems eventually fade or are whittled away to
become dwarf galaxies locally
(e.g. Phillipps \& Driver 1995; Moore et al. 1996).
Previous studies of the lower luminosity population within rich clusters have,
by necessity, concentrated on the most local clusters and groups (e.g.
Impey, Bothun \& Malin 1988; Ferguson \& Sandage 1991; Colless \& Dunn 1996; 
Lobo et al. 1997; Secker 1996; Secker \& Harris 1996; Phillipps et al. 1998; Trentham 1998b) 
and this represents too limited a sample to extract a 
generalised view of the role of dwarf galaxies in rich clusters. However with 
the advent of wide--field imaging cameras with high-efficiency detectors on 
large--aperture telescopes it has now become possible to probe to sufficiently 
low luminosities to sample the dwarf galaxies in more distant clusters (c.f. 
Driver et al. 1994a). This 
photometric method for recovering the cluster luminosity distribution (LD) 
has been tested via exhaustive simulations (Driver et al. 1998; 
hereafter Paper II) and shown to be an effective 
and accurate method for recovering the LDs of {\it rich}
clusters over a broad redshift range $0.1 < z < 0.3$\footnote{Based on our 
chosen detector: AAT f/3.3 + Tek $1024^{2}$ CCD}. A spate of recent
papers from various groups applying this technique find predominantly 
dwarf-rich LDs (e.g. Driver et al. 1994a; De Propris et al. 1995; De Propris
\& Pritchet 1998; Wilson et al. 1996; Smith, Driver \& Phillipps 1997, 
hereafter Paper I; Trentham 1997a,1997b,1998a). On the basis of these results, 
and those from more local clusters (c.f. Godwin \& Peach 1977; Impey, Bothun 
\& Malin 1988; Thompson \& Gregory 1993; Biviano {\it et al} 1995; Lobo et al. 
1997 for example) we postulated the existence of a ubiquitous dwarf rich 
luminosity function for all clusters 
and the possibility of a similar LF for the field (see Driver \& Phillipps 
1996). However, the existing published data represent a highly inhomogeneous 
sample and further work is required to verify or rule out this conclusion.
Here in this third paper of the series, we report the recovered LDs for a 
homogeneous sample of seven rich Abell clusters in the redshift range 
$0.145 < z < 0.168$ which have been observed, reduced and analysed in an
identical manner. The redshift and richness bounds were selected on the
basis of extensive simulations (c.f. Paper II) which suggested that these
criteria were optimal for our chosen detector.
These clusters are: A0022, A0204, A0545, A0868, A2344, 
A2547 and A3888 taken from the catalogue of Abell, Corwin \& Olowin (1989). 
Details of their redshift, distance modulus (here and in what follows we 
adopt $H_{0}=50$\,km\,s$^{-1}$\,Mpc$^{-1}$, $q_{0}=0.5$), richness class, 
and Bautz--Morgan type are given in Table 1.  

The plan of this paper is as follows: 
In \S 2 we describe our observations and the methods used for the reduction 
and calibration of the data. Details of the automated detection and photometry 
of the galaxies within our images are then given in \S 3. In \S 4 the LDs
for each of our clusters are recovered and carefully analysed in terms
of the implied ratio of dwarf to giant galaxies (DGR). In \S 5 we examine the 
variation of this quantity both {\it within} clusters and {\it between}
clusters, revisiting the question as to whether the galaxy LD is dependent
on environment and global cluster properties. We present our conclusions in
\S 6. 

\section{Observations}

Our data were obtained in two separate observing runs carried out in 1996 
at the f/3.3 prime focus of the 3.9\,m Anglo--Australian Telescope (AAT). 
On 1996 January 25, the two clusters, A0545 and A0868, were imaged. This 
night was of excellent quality with photometric conditions and sub--arcsecond  
seeing being experienced throughout the entire night. On 1996 September 14--16,
five further clusters -- A0022, A0204, A2344, A2547 and A3888  -- were imaged 
using a setup identical to that used in the January run. These nights were also
photometric with the seeing varying between 1.0 and 1.3\,arcsec. The mean  
airmass and seeing (image FWHM as measured on our images) over which each  
cluster and its adjacent ``field'' sight--line were observed are listed in 
columns 2 and 3 of Table 2. 

All the data were acquired through the standard Kron--Cousins $R$--band 
using a $1024 \times 1024$ 24\,$\mu$m (0.39\,arcsec) pixel thinned 
Tektronix CCD. This detector gave a total field of view of 
$\sim 6.5 \times 6.5$\,arcmin or 0.0117 sq deg.
Dome flats, twilight flats and bias frames were collected at the start and
end of each night; the dark current of the Tektronix CCD is negligible so 
dark exposures were not taken. The observations of the cluster and adjacent 
``field'' sight-lines were each broken into $9 \times 10$ minute exposures 
with a small 10\,arcsec shift applied between each exposure. This `dithering' 
between exposures was done in a square pattern giving a final overlapping 
field of view of $\sim 6 \times 6$\,arcmin. The field sight-lines were 
selected by simply offsetting the telescope $\sim 75$\,arcmins east of the  
cluster centre. The size of the offset was chosen to be close enough to
sample the 
same large-scale background structure as the cluster while, at the same time, 
avoiding the outer extremities of the cluster itself. To maximise our observing
efficiency and yet ensure that the photometric integrity of our data was not  
compromised, the following sequence of exposures was adopted: 
$8 \times$ cluster$->$standard stars$->$cluster$->$standard stars$->$adjacent field$->$standard stars$->$$8 \times$  adjacent field. The
total exposure time spent on each cluster and its adjacent field region is
listed in column 4 of Table 2.

The standard star exposures in our observing sequence involved the use of 
the photometric sequences set up in Selected Areas 92, 95, 96, 98, 101 and
113 by Landolt (1992) and in E--Regions 2, 8 and 9 by Graham (1982). 
Our observations of each cluster were timed such that 
the airmasses of the standards, field and cluster sight-lines were comparable.  
This obviated the need to make atmospheric extinction corrections and meant 
that we could still work when it was photometric over only short (typically 
30\,min) periods. Changes in transparency were monitored by using the standard
star observations bracketing each cluster/field exposure. 
In eventuality this caution was unnecessary as the measured standards varied 
by no more than $\Delta R = \pm 0.02$\,mag. 
To generate the magnitude zero points for the final frames, zero points were  
first derived for each individual 10\,min cluster and field exposure from its 
bracketing standard star observations.  
By comparing the photometry for a number of objects in the individual field and
the final stacked version, a magnitude zero point was derived
for the entire frame. The advantage of this method is that airmass is 
guaranteed to be taken into account precisely. The residual scatter of the 
standards implied that any colour correction term was small ($\le 0.01$\,mag). 

The initial processing of the data was accomplished in the same way as 
described in Paper I. This involved using the {\sc STARLINK CCDPACK} 
reduction package to
de-bias, flat-field, align and coadd the images. The reduced frames were then  
sky--subtracted (see Driver, Windhorst \& Griffiths 1996) to remove any  
large--scale residual sky gradients -- due mainly to internal scattered light  
from nearby bright stars and/or light leakage. Details of this subtracted
sky background (mean signal in ADU and calibrated surface brightness) and 
its associated noise (in ADU) and 1$\sigma$ surface brightness limit can
be found in columns 5--8 of Table 2. Figure 1 shows a greyscale 
plot of the final coadded and processed images obtained for A0868 and its  
associated field sight--line. Together, Table 2 and Figure 1 indicate that 
the data are of good quality and suitable for the purposes of our LD 
analysis (see Paper II). It is instructive to 
compare these images to the simulated data shown in Paper II, which were used 
to verify the technique we now apply in this paper. 

\section{Image detection and photometry}

\subsection{Generation of catalogues}

The detection and photometry of objects within our CCD images was conducted 
automatically using the SExtractor software package (Bertin \& Arnout 1996) in 
an identical manner to that described in Paper II. For this purpose a 
detection limit of $\mu_{R} = 25.5$\,mag\,arcsec$^{-2}$ over 4 connected 
pixels together with the automated image de-blending algorithm were used; 
the magnitudes measured were Kron (1978) types taken within an aperture of 
radius = $3.5r_{\rm Kron}$. Our measured magnitudes were corrected for Galactic
extinction using values from the study 
of Burstein \& Heiles (1982) and which are available for each cluster on  
NED\footnote{The NASA/IPAC Extra-galactic Database (NED) is 
operated by the Jet Propulsion Laboratory, California Institute of Technology, 
under contract with the National Aeronautics and Space Administration.}; the
actual $A_{R}$ values used are listed in Table 3. 
SExtractor utilises an Artificial Neural Network to perform star/galaxy  
separation and assigns a probability value to each object in the range 0 to 1  
(where 1 indicates a perfectly stellar profile). Figure 2 shows a histogram of the star/galaxy likelihood values obtained for the entire data set.
The peak at 0.95 represents the expected location of stellar objects 
and from this plot we choose to adopt a value of 0.90 as the critical 
division between stars and galaxies. Table 3 summarises the numbers of objects 
classified as stars in each field down to $R\le 23.5$. The shaded area in 
Figure 2 shows the stellar likelihood for those objects with $R\le 20.5$. 

\subsection{The non-cluster sight-lines}

Figure 3 shows the individual galaxy counts (after star-galaxy separation
and reddening correction) for  
each of the seven field sight-lines. The reference line is our optimal
fit to the data of Metcalfe et al. (1995) and is given by
$N(m) = 0.377 - (R_{C} - 12.2)$ over 
$20\le R\le 25$, and $N(m) = 0.457 - (R_{C} -13.56)$ for $R\le 20.0$. 
Incompleteness is indicated by the 
departure of our data from this reference line. 
In all cases other than the B2547 field, there is good qualitative agreement 
between our galaxy counts and those of Metcalfe et al.  
The apparent excess of galaxies in B2547 indicates the possibility of our 
sight-line in this case inadvertently passing through a nearby cluster.  
Examination of the image indicates an obvious excess of bright galaxies across 
this field. This, therefore, precludes us from using B2547 as a reference 
background field.
 
The middle panel of Figure 3 shows
the mean galaxy counts constructed from all our field data (open circles) and
with B2547 excluded (solid squares), minus the fit to the 
Metcalfe et al. data. This shows more quantitatively the good  
agreement between our data and those of Metcalfe
et al. and indicates that our data our complete to $R=23.5$ beyond 
which our counts increasingly fall below the deeper Metcalfe et al. data.
We adopt this value as our completeness limit. Note that
in a number of the published papers on this topic a valiant effort has been 
made to push a little deeper by the application of isophotal correction 
methods. The problem with this approach is that it requires prior knowledge 
of the surface brightness distribution of the faint cluster populations, which 
is unknown. Here we have preferred to take the most conservative course of 
action which is to define our completeness limit as the magnitude at which the
first sign of deviation between our field counts and the much deeper counts of 
Metcalfe et al. appear. In effect we are circumventing surface
brightness selection effects by making a conservatively bright magnitude cut.

The bottom panel of Figure 3 illustrates the percentage variation of observed  
galaxies in each magnitude bin about the mean (as derived from all our 
field sight-lines except B2547). 
The symbols are as defined in the individual number-count plots at the top of
Figure 3. The solid lines represent the anticipated variation in the counts 
from Poisson statistics. At bright magnitudes ($R\le 21.0$) the large variation 
($\sim \pm 50$\%) is a reflection of the limited size of our field of view 
($6 \times 6$ arcminutes). At fainter
magnitudes, down to our completeness limit $R = 23.5$, the variation 
seems slightly in excess of that expected from Poisson statistics. In 
particular, the counts in field B0868 lie systematically below the Poissonian  
limit. There could be several reasons for this: SExtractor incorrectly  
classifying galaxies as stars at the faintest limits, underestimation of the Galactic absorption 
in this relatively low galactic latitude field, or a genuine background inhomogeneity. Cross-referencing the different fields with their observational 
parameters in Table 1 indicates no obvious correlation, in this context, with  
seeing or airmass. The lack of a variation grossly in excess of the Poisson 
limits, on 
the other hand, suggests that it might well be valid to use the mean of all the field sight-lines (except B2547) for the background galaxy subtraction. The
only exception here is at bright magnitudes ($R\le 21$) where it is clear that  
our data cover an insufficient field of view to constitute a fair comparison;  
this is further addressed in \S 4.2.

\subsection{The cluster sight--lines}

The image detection and photometry was performed for the cluster sight-lines
using the same isophotal detection limit and detection parameters as for the 
field sight--lines. The resulting number counts from
the seven cluster sight--lines are shown in Figure 4; again the optimal fit
to the data of Metcalfe et al. (1995) is shown for reference purposes. 
In all cases the counts lie well above the reference line down to the limiting
magnitude of $R=23.5$, indicating the presence of a cluster. The lower
panel shows the percentage excess of galaxies in each magnitude bin over
the mean field counts. Apart from A0204 (our poorest cluster -- see Table 
1), all clusters show a significant excess at all magnitudes down to the 
detection limit. Comparison with Figure 3 indicates that this excess is
significant. 

In looking to establish here whether the form of the LD measured within
each of our clusters is correlated in any way with global cluster properties  
(e.g. richness, BM--class), we have calculated from our data the mean local 
projected galaxy density -- as formulated by Dressler (1980) -- within each 
cluster field. This parameter, which we shall refer to as the Dressler density
parameter (DDP), measures the projected density local to each galaxy by
determining the area (in Mpc$^{2}$) within which its 10 nearest neighbours 
with $M_{V}\le -20.4$\footnote{Equivalent to $R\simeq 19.5$ at the redshift of 
the clusters studied here, assuming $H_{0}=50$\,km\,s$^{-1}$\,Mpc$^{-1}$} 
are contained. Since Dressler showed the morphological (E/S0/Sp) mix of 
galaxies in rich clusters to be a strong and smoothly--varying function of 
this parameter, its consideration within this aspect of the study is clearly
important. The mean DDP value evaluated for each cluster is listed in 
the column 7 of Table 1.   

\section{The luminosity distributions}

\subsection{Construction}

The LD for each cluster was recovered via the statistical subtraction of
the field galaxy counts from those observed towards the cluster. At 
bright magnitudes ($R\le 20.5$) this was accomplished using the  field
counts of Metcalfe et al. (1995). At fainter magnitudes 
($20.5<R\le 23.5$) the counts measured in each cluster's adjacent field
region were used\footnote{Note that for A2547, the nearest acceptable field
region, B0022, was used.}. Star--galaxy separation was performed only
at $R\le 20.5$; at fainter magnitudes it was assumed that each cluster's comparison field sight-line had the same stellar surface density to the cluster 
sight-line. The rationale for using the Metcalfe et al. counts at 
bright magnitudes was because of their greater sky coverage and therefore 
better statistics at bright magnitudes. The use of our adjacent field 
counts at fainter magnitudes was to eliminate or
minimize the following effects: (1)\,surface brightness--dependent 
systematics -- both our cluster and adjacent field catalogues had the same
surface brightness limit, (2)\,subtraction uncertainties due to large--scale
structures, and (3)\,point spread function variations due to changing
atmospheric conditions and/or detector aberrations. 

As part of this process, an adjustment was made for the diminishing field of 
view available to fainter objects (see Paper I for full details). The 
apparent magnitude distributions derived for each cluster were converted to 
ones in absolute magnitude using the distance moduli listed in 
Table 1. To compute the distance moduli a standard flat cosmology 
($\Omega_{o} = 1$, $q_{o} = 0.5$, $H_{0} = 50$ kms$^{-1}$Mpc$^{-1}$) was 
adopted and a uniform $R$--band K-correction of $\Delta R = 1.0 z$ 
(valid for $z < 0.5$; Driver  et al. 1994b) assumed. Figure 5 shows the final  
LDs (solid circles and errorbars) recovered for each of the seven clusters 
and also the mean LD for the entire ensemble.  
The latter was derived from a direct average of the 
individual LDs after normalising each distribution to unity. This ensures that 
the mean LD is not dominated by the richest cluster.
Table 4 tabulates the data plotted in Fig. 5.

Qualitatively, Figure 5 suggests a wide variation in the recovered LDs 
in terms of their faint end ($M_{R}>-21$) behaviour, ranging from the 
flat LD of A3888 to the steep LD of A0868. For reference, the solid line 
in Fig. 5  represents a flat ($\alpha \approx -1.0$) Schechter function 
(Schechter 1976). The larger errorbars for A0204 and A2547 
illustrate the conclusions of Paper II, showing that the LD recovery technique 
eventually fails for poor clusters and this is due to the lack of contrast of
the cluster against the background.  

\subsection{Evaluating the dwarf-to-giant ratios}

Given the quality of the recovered LDs, and the intrinsic restrictions of 
using Schechter function fits (see Paper II), we elect to use a simpler and
more versatile measure to quantify the LD's, viz. 
the ratio of dwarfs ($-19.5 < M_{R} < -16.5$) to giants ($-24.5 < M_{R} < 
-19.5$): 
\noindent
\begin{equation}
\mbox{DGR} = \frac{ \sum \mbox{N}(-19.5 < \mbox{M}_{R} < -16.5) }
{\sum \mbox{N}(-24.5 < \mbox{M}_{R} < -19.5)}.
\end{equation}
This approach was introduced by Ferguson \& Sandage (1991) in their
study of local poor groups and is also favoured by Secker \& Harris (1996) in 
their analysis of the Coma cluster. In column 4 of Table 5 we list the DGR 
values derived from the LDs shown in Figure 5 and we see that they vary from 
0.8 (A3888) to 3.1 (A0204), thus indicating an apparently large range in the  
dwarf component of these rich clusters. For those more familiar with Schechter 
function parameters, this equates to a faint--end slope variation of 
$-0.88 < \alpha < -1.36$; Fig. 6 shows the relation between $\alpha$ and the
DGR parameter if the cluster is 
assumed to be well described by a single Schechter function fit with 
$M_{R}^{*} =-22.5$. It is interesting to note that over bright absolute 
magnitudes ($M_{R} < -20$) all clusters --- with the possible exception of 
A2344 --- exhibit similar LDs confirming the findings of Lumsden et al.
(1997) of the universal nature of the bright end of the cluster LF.
We also note that the cluster LDs do not conform to a strict Schechter
function at bright magnitudes, but show a marginal excess indicative of
the presence of a small population of overly luminous objects in cluster 
environments.

\subsection{Robustness of the background subtraction}

Paper II explored in detail the limitation of our photometric recovery 
technique and illustrated that the reliability of the recovered LD is a strong
function of cluster richness, seeing and redshift, but relatively independent
of LD shape. The critical step in the LD recovery process is the accuracy of
the background subtraction which relies on the contrast of the cluster galaxy
counts against those of the field. The most direct way of testing this
accuracy with real data is to repeat the analysis 
using a different field reference sample. The methodology of the current
recovery is based on the argument that an adjacent field sight-line is the 
most reliable reference and one which mitigates most concerns (accuracy of  
star/galaxy separation, background fluctuations etc). Nevertheless, it is 
arguable from Figure 3 that the variation
in the individual field galaxy counts around the mean is sufficiently small 
that a subtraction based on the mean counts may be equally valid. To 
determine whether this small field--to--field variation has any significant 
effect on the results presented here, 
we reconstruct the LDs using a background subtraction based entirely on the 
mean counts over the full range of apparent magnitude (to compensate for the 
varying galactic latitudes we must now perform star/galaxy classification to 
our magnitude limit). The results are also shown in Figure 5 where they are  
overlaid on top of the original data points (solid circles) and plotted as 
open squares. In most cases the two data sets agree to within
the specified errors. In two cases (A0868 \& A2344) the differences are 
apparently systematic as opposed to random. These two clusters 
both lie close to the Galactic plane and the discrepancy may be a reflection 
of errors in the star-galaxy separation at fainter magnitudes. Despite the  
systematic trend in these two clusters, the broad shape of their LDs is  
consistent, both showing an upturn in their LDs regardless of which background 
subtraction is used. The over-riding qualitative impression from Figure 5 is 
the robustness of the results to the choice of background subtraction. 

To quantify the dependence of our results on the background subtraction we 
recompute the DGR values based on this second reconstruction. These are shown 
in column 5 of Table 5 and can be compared to column 4 which shows the original
values. In all cases the results agree to within their specified errors.

\subsection{How accurately can we measure the DGR?}

This question can be addressed using the mean cluster LD (see Fig. 5) which 
has a DGR of $1.8 \pm 0.2$. Input parameters were assigned to our  
simulation software to produce a simulated f/3.3 AAT image for a cluster at  
z=0.15 with this same DGR value (and mean richness, c.f. Paper II). 
The simulated cluster was then put through the identical detection and 
photometric procedure as the real data and the final DGR value measured as for 
the real data. This procedure was repeated for 400 simulated clusters and 
field sight-lines. The resulting input and output DGR distributions are shown 
in Figure 7. The cross--hatched histogram represents the variation in the input
distribution, with a mean $<$DGR$>$$^{\rm IN}= 1.83$ and standard deviation in 
a single measurement of $\sigma$(DGR)$^{\rm IN} = 0.19$ -- this spread is 
because we only sample the central region of the 
simulated cluster image. The open histogram shows the recovered DGR 
distribution after image simulation, noise addition, detection, star-galaxy 
separation, image de-blending and
final photometry (see Paper II). The recovered distribution is broader than 
the input distribution, as expected due to the process of simulation which
introduces noise and object overlaps, and very slightly skewed to lower DGR 
($<$DGR$>$$^{\rm OUT} = 1.79$ and $\sigma$(DGR)$^{\rm OUT} = 0.29$). 
This series of simulations establishes firmly 
that the recovered DGR is a good representation of the input DGR and that the 
one-sigma measuring accuracy in the DGR is of the order $\pm 0.3$.

\subsection{Verification by simulation}
To directly assess the reliability of the DGR measurements for each cluster, 
we undertook extensive Monte-Carlo simulations based on the method described 
in Paper II. We take as our starting point the parameters listed in Tables 1-5 
and simulate eleven independent cluster and field sight-lines for each of
our actual clusters. From these we derive the recovered DGRs after simulation, 
detection and analysis. Its worth noting that these simulations reflect both 
the statistical variances but also any systematic effects due to disparity 
between the field and cluster seeing and field and cluster noise limits etc. 
Figure 8 shows the resulting histograms of the recovered DGR and the final two 
columns of Table 5 shows the recovered mean DGR and its standard deviation. 
In almost all cases these values are {\it less} than the quoted errorbars, this
is because throughout out error analysis we have assumed all data points in 
the LD are uncorrelated. These simulations underline the robustness of our 
results listed in Table 5. 

It is also interesting to note that two of our clusters (A0204 and A2344) fall 
within the observational domain where reliable results are not guaranteed (see 
Paper II) and we can see that these clusters along with A2547 show the largest 
systematic errors and standard deviation errors in the recovered DGRs. 
Nevertheless these errors, derived via simulations, are smaller than those 
quoted in Column 4 \& 5 of Table 5 and smaller than would be implied by Paper 
II\footnote{The actual GOF values for these simulations are: A0022=84\%, 
A0204=72\%, A0545=91\%, A0868=95\%, A2344=90\% A2547=88\% and A3888=74\%}. 
These is due to a number of reasons. The first is that the DGR is 
a less stringent test than Paper IIs Goodness-of-fit and while the DGR can be 
deemed reliable the exact LD distribution should be treated more cautiously. 
Secondly our actual data has substantially lower noise than that used in the 
earlier simulations of Paper II (our simulations had a 1-$\sigma$ noise limit 
of 26 mags per sq arcsec compared to the numbers listed in column 8 of 
Table 2). The final reason is the general unreliability of the Abell richness
classification (c.f Columns 4 and 7 of Table 1).

\section{Discussion}

As has already been highlighted in the previous section, the LDs we have 
derived and their associated DGRs show considerable variation across our 
sample. This raises a number of issues in relation to the universality of 
the LD, each of which we now address: 

\subsection{Is the data consistent with a ubiquitous cluster LD?}

Having established via simulation the accuracy to which we can measure DGR 
values, we are in a position to assess whether those observed for 
the clusters in our sample are consistent with a universal LD for all cluster  
types. To address this, the distribution of  
observed DGR values is plotted as the solid histogram (scaled up by 
$\times 10$) in Figure 7. The variation in the DGR value for the actual data 
has a mean\footnote{Note that the mean of the observed DGRs and the DGR of the
mean LD are not expected to be identical} of $<$DGR$>$$^{\rm OBS} = 1.9$,
and a standard deviation of $\sigma$(DGR)$^{\rm OBS} = \pm 0.8$ {\it which is  
inconsistent with our seven clusters having a ubiquitous parent LD at the 
$3\sigma$ level}. We therefore consider it unlikely, based on our sample of  
seven clusters, that there exists a ubiquitous cluster LD and conclude that  
clusters are either: `normal' 
($1.0 <$ DGR $< 2.0$), e.g. A0022, A0545, A2547; `dwarf-rich' 
(DGR $> 2.0$), e.g. A0204, A0868, A2547 or `dwarf-poor' (DGR $< 1.0$), 
e.g. A3888. 

\subsection{Is the DGR uniform throughout the cluster?}

In Paper I we identified a luminosity--dependent segregation in A2554 in 
that the giant galaxies were more centrally concentrated than the dwarfs. If
this is universal then we expect the DGRs to increase as a function of radius.
Unfortunately the new data have a more limited coverage (typically 0.74  
Mpc radius from cluster centre) than the A2554 data (1.40 Mpc), hence a full 
analysis of this segregation requires further data.
However, we can make a preliminary investigation by 
measuring the DGRs for ``inner'' ($r\le 0.56$\,Mpc) and ``outer'' 
($0.56<r\le 0.74$\,Mpc) regions of the clusters observed here; 
Table 5 shows the results. For clusters A0022 and A3888 the DGR values
decrease in the outer regions contrary to the trend seen in A2554. For all
other clusters the DGR increases significantly from the inner to outer region
corroborating the trend seen in A2554. {\it This further emphasises the fact  
that the LD and DGR observed within a cluster depends sensitively on over 
what physical region of the cluster the measurement is made.} In this respect
our data reveal two important trends: (1)\,significant variations in the DGR 
even over the limited radial limits ($r\ls 1.2$\,Mpc) probed here, but also
importantly (2)\,variations in the DGR from cluster to cluster are at a
minimum within the central $\sim 0.5$\,Mpc. Clearly further data need to be 
acquired to firm up these two results and these will be reported in future 
papers in this series.     

\subsection{Possible correlations between the DGR and cluster properties}

Drawing upon the above results, Figure 9 shows the DGR values measured for the 
inner regions (top), the outer regions (middle), and the entire cluster 
(bottom), plotted against Bautz-Morgan class (Bautz \& Morgan 1970), Abell 
richness and the mean DDP values derived for these regions. 
As noted in \S 5.2, there is less
variation from cluster to cluster in the DGR observed in the inner
regions and we see this graphically in the top panels of Figure 9 where 
the DGR values are mostly in agreement to within their uncertainties.
{\it For some reason, the balance between dwarf and giant galaxies appears to
be similar within the inner cores ($r\le 0.5$\,Mpc) of clusters
irrespective of their global properties.} Taken at face value, this suggests 
some physical mechanism operating in the core of clusters between the giants 
and dwarfs. However, it is important to remember that it has not been 
established whether the dwarfs are actually in the core or are simply the 
projected density of a surrounding halo along the line-of-sight through the 
core. 

In contrast to the inner regions, the outer regions do show trends, 
with the DGR exhibiting a reasonable correlation with BM--class and a strong
anti--correlation with mean DDP. This is also reflected weakly in the 
``total'' cluster plots shown in the bottom panels. In these outer 
regions, therefore, it is quite clear that either the dwarf galaxy population 
is diminished or the giant galaxy population is boosted in clusters where 
there is a higher overall local density of galaxies and which are dynamically 
more evolved (i.e. lower BM--class). The diminution of the dwarf population in
this context is certainly consistent with the ``galaxy harassment'' scenario 
(Moore et al. 1996) which would predict that this process is most effective
and has been in operation longer in the more relaxed and centrally dense  
clusters, thereby significantly whittling away their lower luminosity 
populations and hence reducing their DGR. 

\subsection{Is there a radial dependence? -- A2554 revisited}

All our new cluster LDs have substantially 
flatter faint--end slopes than that reported in recent publications, 
including our own (Driver et al. 1994a; 
Paper I; Wilson et al. 1997). Both the Driver et al. and Wilson 
et al. studies relied on isophotal corrections which are highly 
dependent on the assumed profiles for the fainter cluster members 
(Trentham 1998b). However, the more recent study of A2554 (Paper I) 
relied on no such correction which begs the question as to why A2554 shows 
a significantly higher DGR (DGR$_{A2554} = 4.2 \pm 0.3$)? One possible reason, 
mentioned in \S 4.6, is the significantly larger areal coverage of the A2554  
data ($r\sim1.40$\,Mpc cf $r < 0.74$\,Mpc for the seven clusters of this 
study). If the fainter galaxies
{\it are} distributed more uniformally than the giants one might expect the DGR
to increase with radius. To test this we reconstruct the DGR for A2554 as a
function of radius. The results are shown in Figure 10 with the DGR being
plotted both differentially (top panel) and cumulatively (bottom panel) as
a function of radius. The vertical dashed line indicates the radial limit
applicable to the 7 new clusters presented in this study. Both the 
cumulative and differential DGR show a marked increase as a function of 
radius, implying that for A2554 the dwarf population is more broadly 
distributed than the giants (see also Paper I). At a radius
of 1\,Mpc from the cluster centre the local DGR value is approximately 6-8
(comparable with a single Schechter function slope of $\alpha \approx -1.7$)
as opposed to the central value of 3 ($\alpha \approx -1.3$).

\subsection{Decomposing the DGRs}
Up until now we have considered only the DGR and have not explicitly examined
the absolute numbers of giants and dwarfs within our clusters. In Figure 11,
therefore, we plot the actual density of giants and dwarfs per Mpc$^{2}$ for
the inner and outer regions separately. For clarity the inner and outer giant 
points are linked by solid lines while the dwarf points are linked by dashed 
lines. The clusters are arranged from top to bottom in order of decreasing DDP 
with their Bautz-Morgan class denoted alongside. For the 
two densest clusters (A3888 \& A0022) the dwarf galaxies actually exhibit a 
marginally steeper radial drop-off than for the giants. These two clusters are 
also low Bautz-Morgan types where spherical symmetry is more likely to be an 
accurate representation of the galaxy distribution and they are generally 
more relaxed. An additional concern is lensing. However Trentham (1998a) 
showed that lensing in the core region of a very rich cluster will enhance the 
background density by only $\sim 10$\%.
For the remaining clusters, the dwarfs exhibit a significantly 
flatter radial drop-off and in three cases, A0545, A2547 \& A0204, actually
{\it increases} from the cluster centre -- determined, of course, from the 
giant population. It is very difficult to understand how the density of dwarfs 
could possibly increase with respect to the inner region if the 
dwarfs exist in a smoothly distributed spherical profile around the primary 
cluster center\footnote{Since both the near-side and far-side outer dwarf 
populations would project onto the ``inner'' sight-line.}. One intriguing 
possibility is that the core regions are devoid of dwarf galaxies, although 
why this should be the case for the poor rather than rich clusters is not 
clear. However, we note that these three clusters are all Bautz-Morgan 
class III and a second explanation is that in such clusters the dwarfs 
exhibit extremely asymmetrical profiles as might be 
expected in these non-relaxed environments. This asymmetry combined with a 
higher DGR in lower density environments would be sufficient to explain the 
observed increase.

\section{Conclusions}
We have recovered the luminosity distributions (from $-24.5\le M_{R}\le -16.5$)
within the central regions 
($r\le 0.74$\,Mpc) of seven clusters at $z\simeq 0.15$. We 
find a variation in the recovered DGRs of 0.8---3.1 (equivalent 
to: $-0.9\le \alpha\le -1.4$). This variation is found to be inconsistent at 
the 3$\sigma$ level with the notion of a universal cluster LD. However,  
we find that the LDs derived for the inner ($r\le 0.56$\,Mpc) core regions 
show more universality than those derived for the exterior ($r> 0.56$ Mpc) 
regions. The implication is that the clusters studied here show 
a wide range in the richness of their dwarf halos from very poor 
(DGR $ \sim 0.7$) to extremely dwarf--rich (DGR $\sim 7.6$), equivalent to 
$\alpha = -0.8$ and $-1.7$ respectively. Furthermore, there is a strong 
anti-correlation between the DGR and the mean local projected (Dressler) 
density in that clusters with lower overall projected densities contain a 
greater relative density of dwarf systems. 

Re-analysis of earlier data on A2554 (see Paper I) re-confirms the 
strongly clustered nature of giant galaxies ($M_{R}\le -19.5$),
as opposed to the dwarf systems ($M_{R} > -19.5$) for this cluster.
Examining the densities of dwarf and giant galaxies 
independently in the inner and outer regions for our latest clusters 
suggests that for the densest and most relaxed clusters (i.e. 
Bautz-Morgan class I) the dwarfs, while seen in relatively fewer numbers,
follow a similar density drop-off to the giants. Conversely, the lower density,
dynamically unevolved clusters (i.e. Bautz-Morgan class III) exhibit an increase
in their dwarf densities in the outer regions, suggesting that the dwarf-rich 
halos in these clusters are highly asymmetrical and extend well
beyond the cluster radius surveyed (r$\ls 0.74$\,Mpc as seen for A2554).
This result, in conjunction with the general observed trend of higher relative
density of dwarf to giant systems in less dense environments, implies a high 
density of dwarf systems in the field (see Marzke et al. 1994; Driver \& 
Phillipps 1996).

The broad conclusion is that the distribution of dwarf galaxies does not 
necessarily follow that of the more luminous cluster members particularly for 
high Bautz-Morgan class clusters, implying an initial luminosity segregation 
and, more importantly, that the dwarfs have trod a separate evolutionary path 
than that of the giants. This segregation has already been predicted from the 
numerical models of Kauffmann et al. (1997). 
As we have previously discussed (\S 4.6), 
galaxy ``harassment'' may be a prime cause of this luminosity segregation due
to the lower luminosity systems being the most susceptible to the strong tidal
forces they encounter when entering a cluster (Moore et al. 1996). Further
work is required -- in particular, deep wide-field optical imaging -- to 
determine how far the halo of dwarf galaxies around clusters extends and to  
quantify its possible contribution to the inferred dark matter halos of 
clusters (Navarro, Frenk \& White 1996). Probing to higher redshifts will 
also establish whether the trends seen in the current sample evolve and may  
provide strong constraints to the evolutionary processes at work in rich  
clusters.

Speculatively then, the results presented here are consistent with a scenario
in which clusters start from initially extended asymmetrical dwarf-rich 
environments and evolve towards fully relaxed giant rich clusters with 
severely diminished dwarf-poor halos.
We note that remarkably similar results to those presented here have been 
found by L\^{o}pez-Cruz (1997) based on an entirely independent study of
northern Abell/ACO clusters.

\section*{Acknowledgments}
The results presented here are based on observations made with the AAT 
and we thank the support and technical staff of the Anglo-Australian 
Observatory for their valuable assistance. 
We thank Stuart Ryder for riding the prime focus cage at the AAT during the
January observations. SPD and WJC acknowledge the financial support of the Australian 
Research Council throughout this work. SP is supported by the Royal Society
via a University Research Fellowship.

\section*{References}

\begin{description}
\item Abell G., 1958, ApJS, 3, 1.

\item Abell G., Corwin H.G., Olowin R., 1989, ApJS, 301, 83.

\item Abraham R.G., Tanvir N.R., Santiago B.X., Ellis R.S., Glazebrook K.,
van den Bergh S., 1996, MNRAS, L47

\item Barger A., Arag\'{o}n-Salamanca A.S., Smail I., Ellis R.S., 
Couch W.J., Butcher H., Dressler A., Oemler A., Poggianti B.M., Sharples R.M., 
1998, ApJ, in press

\item Bautz L.P., Morgan W.W., 1970; ApJ, 162, L149

\item Bernstein G.M., Nichol R.C., Tyson J.A., Ulmer M.P., Wittman D., 1995 
AJ, 110, 1507

\item Bertin E., Arnout S., 1996, A\&AS, 117, 393

\item Biviano A., Durret F., Gerbal D., Le F\`{e}vre O., Lobo C., Mazure A., 
Slezak E., 1995, A\&A, 297, 610

\item Burstein D., Heiles C., 1984, ApJS, 54, 33

\item Butcher H., Oemler A. Jr., 1978, ApJ, 219, 18

\item Charlot S., Silk J., 1994, 432, 453

\item Colless M., 1995, in {\it Wide Field Spectroscopy and the Distant 
Universe}, ed. S.J.Maddox \& A.Aragon-Salamanca (Singapore: World Scientific), 
263

\item Colless M., Dunn A.M., 1996, ApJ, 458, 435

\item Couch W.J., Ellis R.S., Sharples R.M., Smail I., 1994, ApJ, 430,121

\item Couch W.J., Barger A.J., Smail I., Ellis R.S., Sharples R.M., 1998, ApJ, 
497, 188

\item De Propris, R., Pritchet, C.J., Harris, W.E., McClure, R.D., 1995, ApJ, 
450, 534

\item De Propris, R., Pritchet, C.J., 1998, ApJ, submitted (astro-ph/9805281)

\item Dressler A., 1978, ApJ, 223, 765

\item Dressler A., 1980, ApJ, 236, 351

\item Dressler A., Oemler A. Jr., Butcher H.R., Gunn J.E., 1994, ApJ, 435, 
23

\item Driver S.P., Phillipps S., Davies J.I., Morgan I., Disney M.J., 1994a, 
MNRAS, 268, 393.

\item Driver S.P., Phillipps S., Davies J.I., Morgan I., Disney M.J., 1994b, 
MNRAS, 266, 155

\item Driver S.P., Windhorst R.A., Griffiths R.E., 1995, ApJ, 453, 48

\item Driver S.P., Windhorst R.A., Ostrander E.J., Keel W.C., Griffiths R.E.,
Ratnatunga K.U., 1995, ApJL, 449, L23

\item Driver S.P., Phillipps, S., 1996, ApJ, 469, 529 

\item Driver S.P., Couch W.J., Phillipps S., Windhorst R.A., 1996, ApJ, 466, L5

\item Driver S.P., Couch W.J., Phillipps S., Smith R.M., 1998, MNRAS, in press
(Paper II)

\item Driver S.P., Couch W.J., Odewahn S.C., Windhorst R.A., 1997,
in proc 18$^{th}$ Texas symp. on ``Relativistic Astrophysics'', in press

\item Ellis R.S., Colless M., Broadhurst T.J., Heyl J., Glazebrook K., 1996, 
MNRAS, 280, 235

\item Ellis R.S., 1997, ARA\&A, 35, 389

\item Ferguson H., Sandage A., 1991, AJ, 96, 1620

\item Godwin J., Peach J.V., 1977, MNRAS, 181, 323.

\item Graham J.A., 1982, PASP, 94, 244

\item Impey C., Bothun G.D., Malin D., 1988, ApJ, 330, 634

\item Kauffmann G., Nusser A., Steinmetz M., 1977, MNRAS, 286, 795

\item Koo D.C., Kron R.G., 1992, ARA\&A, 30, 613

\item Kron R.G., 1978, PhD, Univ. Berkeley

\item Landolt A.U., 1992, AJ, 104, 1

\item Lilly S., Tresse L., Hammer F., Crampton D., Le F\`{e}vre O., 1995, ApJ, 
455, 108

\item Lobo C., Biviano A., Durret F., Gerbal D., Le F\`{e}vre O., Mazure A.,
Slezak E., 1997, A\&A, 317, 385

\item L\^{o}pez-Cruz O., 1997, PhD, Univ. Toronto

\item L\^{o}pez-Cruz O., Yee H.K.C., Brown J.P., Jones C., Forman W., 1997, 
ApJL, 475, 97

\item Lumsden S., Collins C.A., Nichol R.C., Eke V.R., Guzzo L., 1997, 
MNRAS, 290, 119

\item Marzke R., Huchra J.P., Geller M.J., 1994, ApJ, 428, 43

\item Metcalfe N., Shanks T., Fong R., Roche N., 1995, MNRAS, 273, 257 

\item Moore B., Katz N., Lake G., Dressler A., Oemler A Jr., 1996, Nature, 379,
613

\item Moore B., Lake G., Katz N., 1998, ApJ, 495, 139 

\item Navarro J.F., Frenk C.S., White S.D.M., 1996, ApJ, 462, 563

\item Oemler A. Jr., 1974, ApJ, 194, 1

\item Oemler A. Jr., Dressler A., Butcher H.R., 1997, ApJ, 474, 561

\item Phillipps S., Driver S.P., 1995, MNRAS, 274, 832

\item Phillipps S., Parker Q.A., Schwartzenberg J.M., Jones J.B., 1998, ApJ, 493, L59

\item Schade D.L., Barrientos F., L\^{o}pez-Cruz O., 1997, 477, L17

\item Schechter P., 1976, ApJ, 203, 297

\item Secker J., 1996, ApJ, 469, 81

\item Secker J., Harris W.E., 1996, ApJ, 469, 623

\item Smail, I. Dressler A., Couch W.J., Ellis R.S., Oemler A.Jr., Butcher H.,
Sharples R.M., 1997, ApJS, 110, 213

\item Smith R.M., Driver S.P., Phillipps S., 1997, MNRAS, 287, 415 (Paper I)

\item Steidel C.C., Dickinson M., Persson S.E., 1994, ApJ, 437, L75

\item Thompson L.A., Gregory S.A., 1993, AJ, 106, 2197 

\item Trentham N., 1997a, MNRAS, 286, 133

\item Trentham N., 1997b, MNRAS, 290, 334

\item Trentham N., 1998a, MNRAS, 295, 360

\item Trentham N., 1998b, MNRAS, 293, 71

\item Tr\`{e}vese D., Crimele G., Appodia B., 1996, A \& A, 315, 365

\item Wilson G., Smail I., Ellis R.S., Couch W.J., 1997, MNRAS, 284, 915

\end{description}

\pagebreak

\section*{Tables}

\begin{table}[h]
\begin{center}
\caption{Cluster sample and properties}
\begin{tabular}{ccccccccc}\hline \hline
Cluster &$z$&$D_{mod}^{1}$&Abell&BM-type&FOV&DDP \\ 
        &   &             & Richness &           &(Mpc$^{2}$) &
(gals\,Mpc$^{-2}$) \\ \hline
A0022 & 0.143 & 39.85 & 3 & I   & 1.53 & $71.9$ \\
A0204 & 0.156 & 40.09 & 1 &III  & 1.75 & $13.9$ \\
A0545 & 0.154 & 40.06 & 4 &III  & 1.72 & $69.3$ \\ 
A0868 & 0.153 & 40.04 & 3 &II-III&1.70 & $49.6$ \\
A2344 & 0.145 & 39.91 & 1 &II   & 1.56 & $50.6$ \\
A2547 & 0.149 & 39.98 & 2 &III  & 1.64 & $32.8$ \\
A3888 & 0.168 & 40.27 & 2 &I-II & 1.96 & $82.3$ \\
\hline
\end{tabular}
\end{center}
\noindent
$^{1}$ Assuming H$_{o}=50$ kms$^{-1}$ Mpc$^{-1}$ and $q_{o}=0.5$.
\end{table}

\begin{table}[h]
\begin{center}
\caption{Characteristics of our AAT f/3.3 imaging data}
\begin{tabular}{lccccccc} \hline \hline
Field$^{1}$&Mean airmass&FWHM&Exp&Background&$\mu_{R}(Sky)$&Noise & 
1$\sigma$SBL$^{3}$ \\
           &            &(arcsec)&(min)&(ADU)&(mag\,arcsec$^{-2}$)&(ADU) & 
(mag\,arcsec$^{-2}$)\\ \hline
A0022&1.06&1.26&100&2578&21.0&12&26.8 \\
B0022&1.10&1.35&100&3197&20.7&14&26.6 \\
A0204&1.12&1.07&100&2689&21.0&13&26.8 \\
B0204&1.11&1.25&90&2423&20.9&11&26.8 \\
A0545&1.08&1.02&90&3266&20.5$^{2}$&21&25.9 \\
B0545&1.10&1.04&90&2234&20.9&13&26.5 \\
A0868&1.10&1.02&90&2523&20.7&13&26.5 \\
B0868&1.13&1.02&90&2317&20.8&14&26.4 \\ 
A2344&1.04&1.15&70&2732&20.9&19&26.3 \\
B2344&1.03&0.95&100&2726&20.9&10&27.0 \\
A2547&1.15&1.10&90&2453&20.9&17&26.3 \\
B2547&1.02&1.05&90&2216&21.0&9&27.0 \\
A3888&1.10&1.27&100&2987&21.1&19&26.3 \\
B3888&1.04&1.03&90&2338&21.1&14&26.7 \\
\hline
\end{tabular}
\end{center}
\noindent
$^{1}$ A denotes the Abell cluster number and B the corresponding nearby field 
sight-line.

\noindent
$^{2}$ The brighter sky for A0545 is due to the setting moon resulting in a 
higher mean background and higher noise.

\noindent
${^3}$ 1$\sigma$SBL:- 1 $\sigma$ surface brightness limit

\end{table}

\vspace{1.0cm}

\begin{table}[h]
\begin{center}
\caption{Stellar density}
\begin{tabular}{crrrrr} \hline \hline
Cluster & \multicolumn{2}{c}{Stars} & l & b & $A_{R}$\\
ID & Cluster & Field \\ \hline
A0022 &  56 &  63 &  42.89 & -82.98 & 0.04 \\
A0204 &  58 &  43 & 148.09 & -67.83 & 0.06 \\
A0545 & 109 & 145 & 214.60 & -22.72 & 0.20 \\
A0868 & 138 & 116 & 244.72 & +32.49 & 0.03 \\
A2344 & 151 & 153 &  29.02 & -42.87 & 0.05 \\
A2547 &  48 &  63 &  42.14 & -66.13 & 0.03 \\
A3888 &  93 &  70 &   3.97 & -59.40 & 0.00 \\ \hline
\end{tabular}
\end{center}
\end{table}

\small

\hspace{-1.0cm}
\begin{table}[h]
\caption{Tabulated luminosity distributions for the seven clusters and their 
mean}
\hspace{-1.0cm}
\begin{tabular}{rrrrrrrrrrr} \hline \hline
M$_{R}$ &A0022    &A0204    &A0545    &A0868     &A2344    &A2547    &A3888    &A963      &A2554     &MEAN\\ \hline
$-24.75$&$ 0\pm 0$&$ 0\pm 0$&$ 1\pm 1$&$  0\pm 0$&$ 1\pm 0$&$ 0\pm 0$&$ 0\pm 0$&$  0\pm 0$&$  2\pm 1$&$ 0\pm 0$\\
$-24.25$&$ 1\pm 1$&$ 0\pm 0$&$ 1\pm 1$&$  1\pm 1$&$ 0\pm 0$&$ 0\pm 0$&$ 4\pm 2$&$  0\pm 0$&$  1\pm 1$&$ 1\pm 0$\\
$-23.75$&$ 3\pm 2$&$ 2\pm 1$&$ 2\pm 1$&$  0\pm 0$&$ 2\pm 1$&$ 1\pm 1$&$ 3\pm 2$&$  0\pm 1$&$  2\pm 1$&$ 2\pm 1$\\
$-23.25$&$ 2\pm 1$&$ 1\pm 1$&$ 3\pm 2$&$  4\pm 2$&$ 0\pm 0$&$ 2\pm 1$&$ 8\pm 3$&$  1\pm 1$&$  2\pm 1$&$ 3\pm 1$\\
$-22.75$&$ 8\pm 3$&$ 4\pm 2$&$11\pm 4$&$  5\pm 2$&$ 2\pm 2$&$ 4\pm 2$&$13\pm 4$&$  3\pm 2$&$  6\pm 2$&$ 7\pm 1$\\
$-22.25$&$ 5\pm 3$&$ 5\pm 3$&$17\pm 4$&$ 11\pm 4$&$ 5\pm 3$&$ 9\pm 3$&$22\pm 5$&$  7\pm 3$&$ 15\pm 4$&$11\pm 2$\\
$-21.75$&$17\pm 4$&$ 2\pm 2$&$20\pm 5$&$ 18\pm 5$&$14\pm 4$&$ 8\pm 3$&$24\pm 5$&$ 10\pm 3$&$ 20\pm 5$&$16\pm 2$\\
$-21.25$&$25\pm 5$&$ 5\pm 3$&$24\pm 5$&$ 20\pm 5$&$11\pm 4$&$ 7\pm 3$&$26\pm 6$&$ 16\pm 4$&$ 27\pm 5$&$18\pm 2$\\
$-20.75$&$16\pm 5$&$ 8\pm 4$&$30\pm 6$&$ 21\pm 5$&$14\pm 5$&$20\pm 5$&$38\pm 7$&$ 18\pm 4$&$ 33\pm 6$&$23\pm 2$\\
$-20.25$&$17\pm 5$&$17\pm 6$&$30\pm 7$&$ 36\pm 7$&$16\pm 5$&$18\pm 6$&$35\pm 8$&$ 18\pm 4$&$ 41\pm 6$&$25\pm 3$\\
$-19.75$&$21\pm 7$&$22\pm 7$&$42\pm 8$&$ 26\pm 7$&$29\pm 7$&$14\pm 6$&$29\pm 8$&$ 16\pm 4$&$ 55\pm 7$&$27\pm 3$\\
$-19.25$&$11\pm 8$&$37\pm 9$&$37\pm10$&$ 42\pm10$&$24\pm11$&$13\pm 8$&$27\pm10$&$ 18\pm 4$&$ 63\pm 8$&$27\pm 4$\\
$-18.75$&$16\pm 9$&$ 7\pm10$&$ 9\pm11$&$ 57\pm12$&$31\pm12$&$32\pm10$&$24\pm11$&$ 32\pm 6$&$ 90\pm 9$&$28\pm 5$\\
$-18.25$&$24\pm11$&$40\pm11$&$90\pm14$&$ 50\pm13$&$33\pm12$&$34\pm12$&$43\pm14$&$ 47\pm 7$&$110\pm11$&$46\pm 6$\\
$-17.75$&$18\pm14$&$21\pm13$&$52\pm15$&$ 78\pm14$&$47\pm15$&$23\pm14$&$51\pm16$&$ 68\pm 8$&$140\pm12$&$41\pm 7$\\
$-17.25$&$30\pm16$&$38\pm17$&$52\pm17$&$ 73\pm17$&$39\pm18$&$ 0\pm15$&$11\pm19$&$ 88\pm 9$&$198\pm14$&$32\pm 8$\\
$-16.75$&$37\pm19$&$62\pm20$&$76\pm18$&$112\pm20$&$45\pm20$&$19\pm18$&$ 0\pm20$&$114\pm11$&$259\pm16$&$47\pm 9$\\
$-16.25$&$46\pm21$&$78\pm22$&$85\pm19$&$134\pm23$&$52\pm23$&$11\pm21$&$ 0\pm21$&$  NA $&$  NA$&$53\pm10$\\ \hline
\end{tabular}
\end{table}

\begin{table}[h]
\begin{center}
\caption{Dwarf-to-giant ratios for the cluster sample.}
\begin{tabular}{cccccccc} \hline \hline
Cluster & Giants & Dwarfs &\multicolumn{5}{c}{Dwarf-to-Giant ratios} \\ 
Name    &\footnotesize{$-24.5<M_{R}<-19.5$}&\footnotesize{$-19.5<M_{R}<-16.5$}& Final & Alternate & $r<0.56$Mpc & $r>0.56$Mpc & Simulated \\ \hline\hline
A0022 & $114\pm13$ & $135\pm32$ & $1.2\pm0.3$ & $1.4\pm0.3$ & $1.3\pm0.3$ & $0.9\pm1.5$ & $1.2\pm0.2$ \\    
A0204 & $ 67\pm12$ & $205\pm34$ & $3.1\pm0.7$ & $2.7\pm0.6$ & $1.3\pm0.5$ & $7.4\pm3.0$ & $3.3\pm0.3$ \\    
A0545 & $180\pm16$ & $316\pm36$ & $1.8\pm0.3$ & $1.8\pm0.3$ & $1.3\pm0.3$ & $2.4\pm0.5$ & $1.7\pm0.1$ \\    
A0868 & $141\pm14$ & $411\pm36$ & $2.9\pm0.4$ & $2.2\pm0.3$ & $2.7\pm0.5$ & $3.3\pm0.7$ & $3.0\pm0.2$ \\    
A2344 & $ 93\pm12$ & $219\pm37$ & $2.3\pm0.5$ & $3.0\pm0.6$ & $2.1\pm0.4$ & $3.8\pm2.6$ & $2.3\pm0.3$ \\     
A2547 & $ 83\pm12$ & $121\pm33$ & $1.5\pm0.5$ & $1.8\pm0.5$ & $1.0\pm0.4$ & $3.1\pm1.6$ & $1.2\pm0.3$ \\     
A3888 & $201\pm17$ & $157\pm38$ & $0.8\pm0.2$ & $1.0\pm0.3$ & $0.9\pm0.2$ & $0.7\pm0.4$ & $0.7\pm0.1$ \\ \hline
\end{tabular}
\end{center}
\end{table}

\normalsize
 
\pagebreak

\begin{figure}[p]
 
\vspace{-3.0cm}

\centerline{\psfig{file=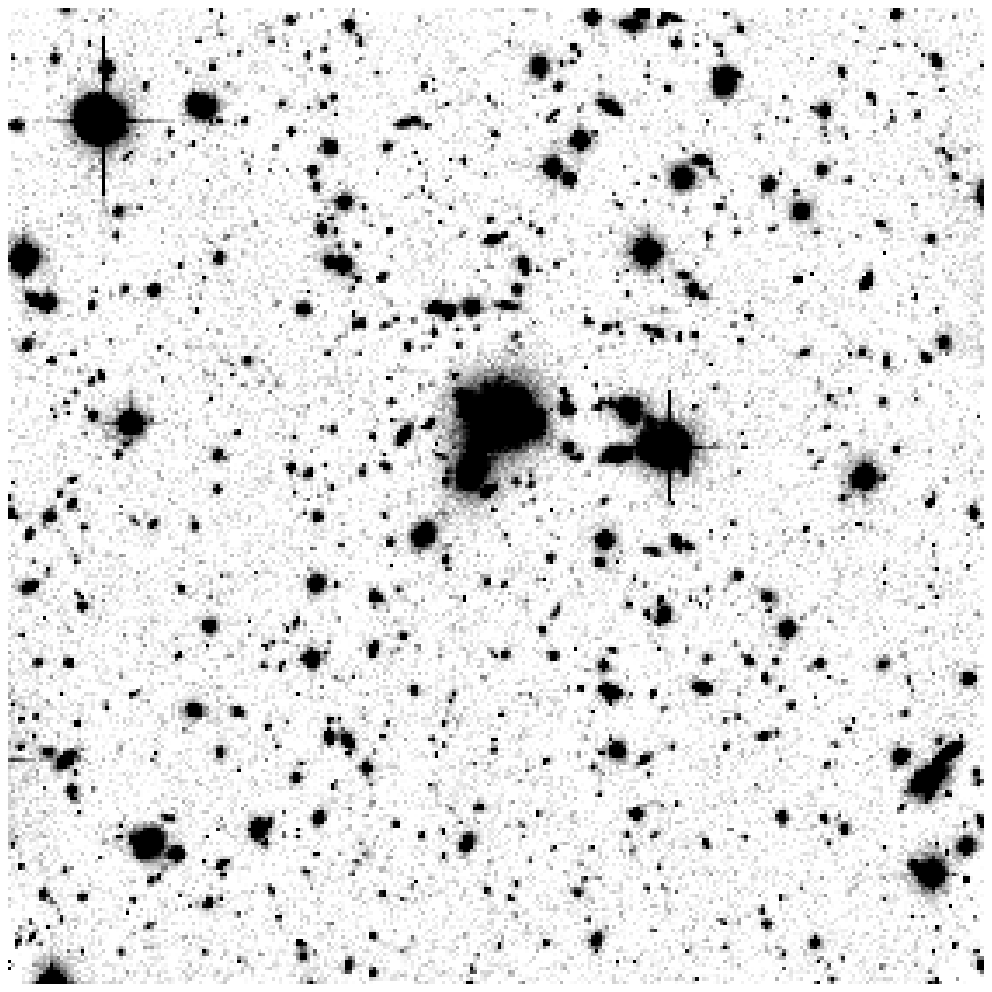}}

\end{figure}

\begin{figure}[p]

\vspace{-3.0cm}

\centerline{\psfig{file=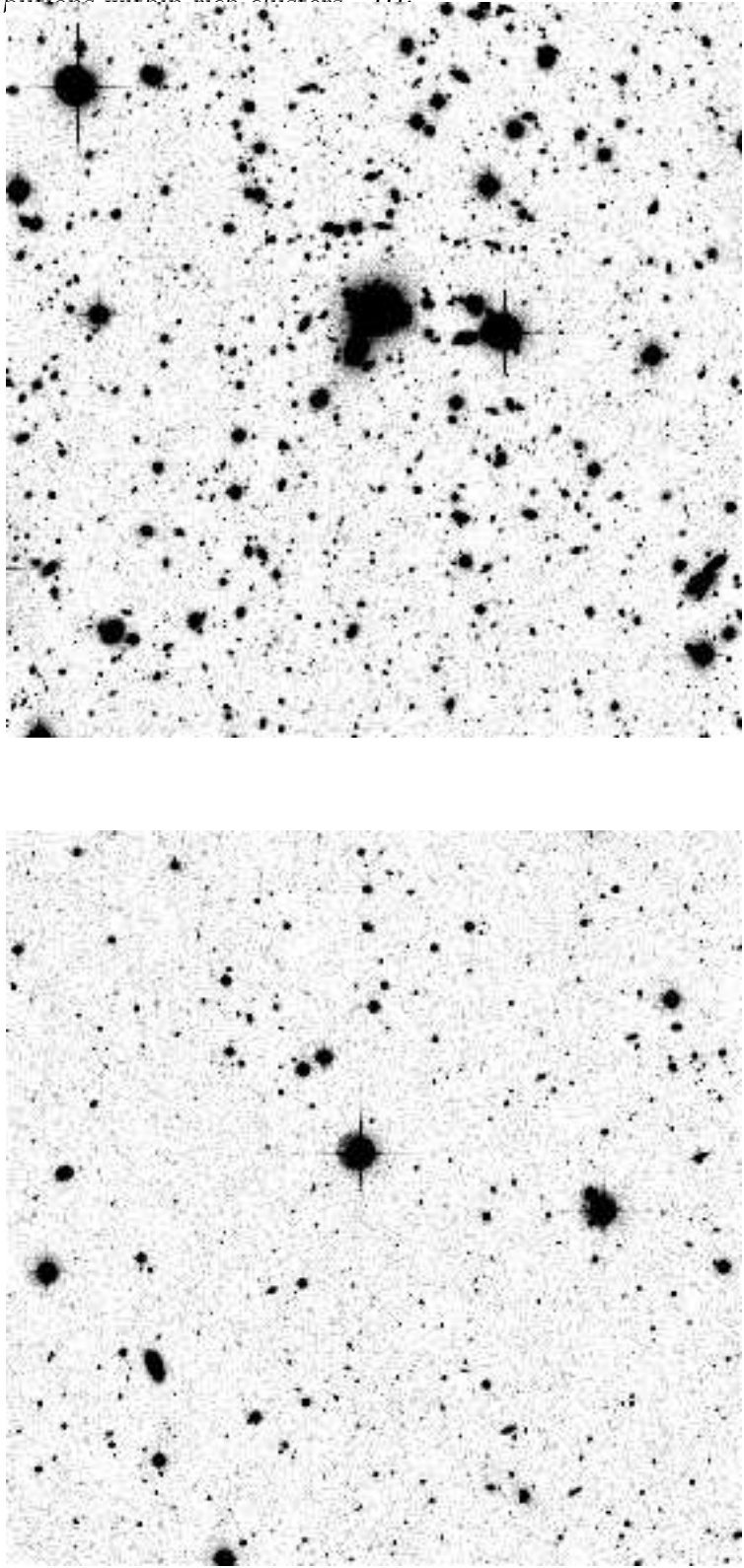}}

\caption{A greyscale plot of one of our cluster sight-lines (A0868; top) and 
its associated nearby field sight-line (B0868; bottom).}
\end{figure}

\begin{figure}[p]
\centerline{{\psfig{file=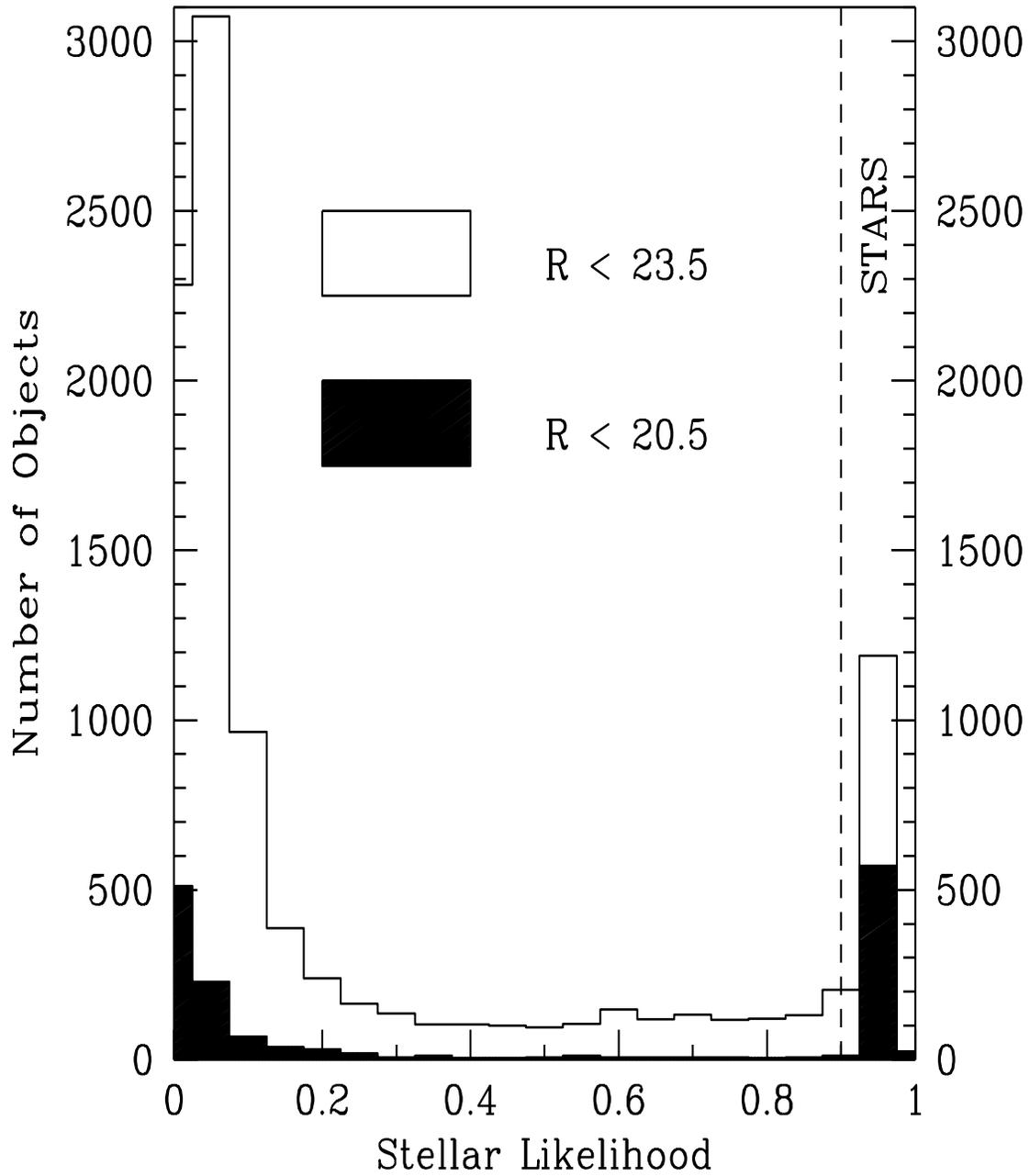,height=200mm,width=160mm}}}
\caption{Histogram of the star/galaxy likelihood parameter. The open 
histogram is for objects with $R\le 23.5$ and
the shaded region for $R\le 20.5$. A value of 0.9 is adopted as the 
critical distinction between stars and galaxies.}
\end{figure}

\begin{figure}[p]
\vspace{-0.5cm}
\centerline{{\psfig{file=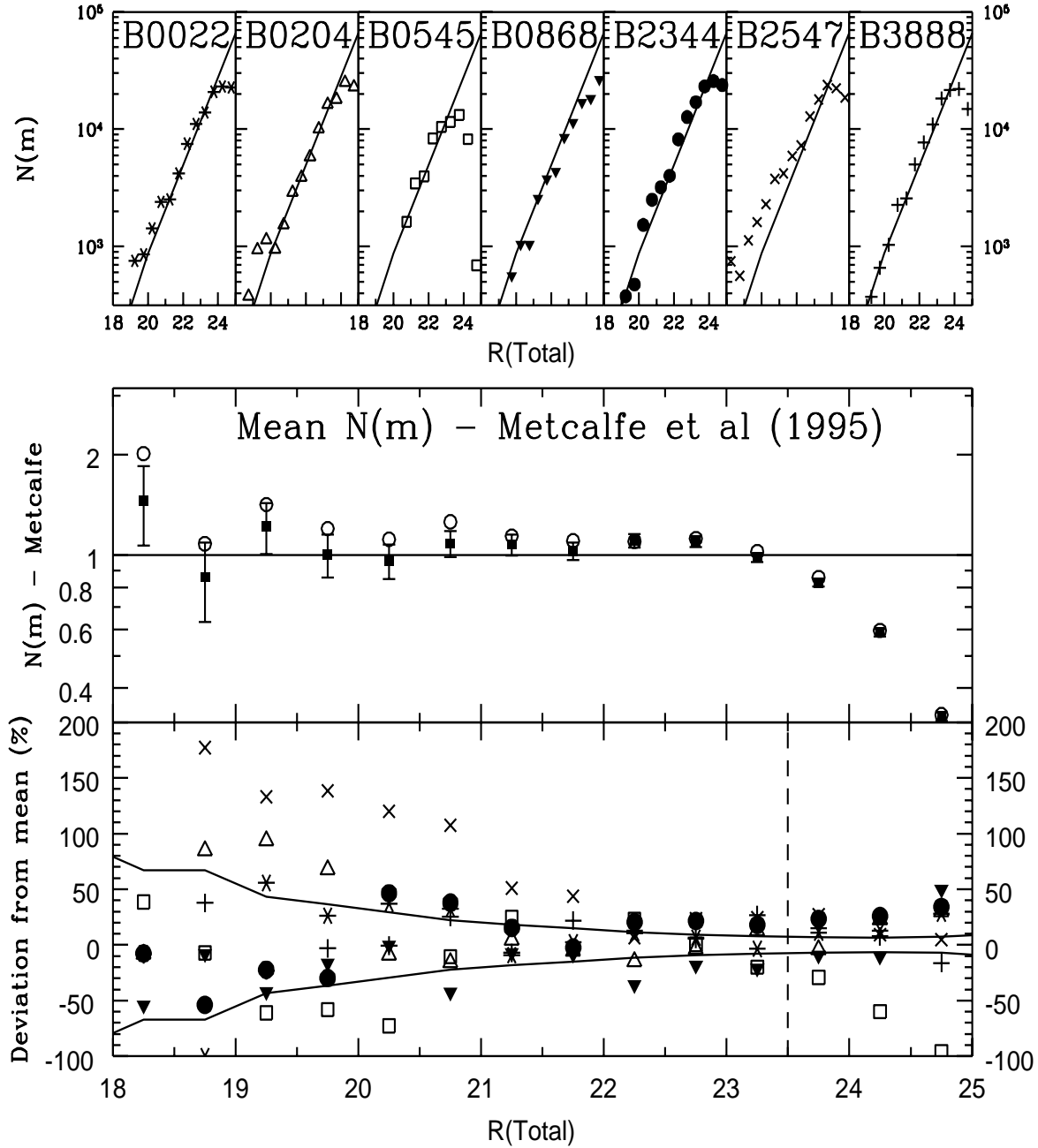,height=180mm,width=160mm}}}
\caption{The top seven panels show the galaxy number-counts for the seven
nearby field sight-lines (as indicated). Error-bars are not shown for clarity.
The reference line is an optimal fit to the published counts of Metcalfe 
et al. (1995). Note that while six of the field sight-lines show close 
agreement, B2547 shows a distinct excess indicating the possible presence of a 
nearby cluster. The middle panel shows the mean galaxy counts derived from 
all the field sight-lines 
(solid circles) and from all fields except B2547 (squares 
and errorbars). Our combined galaxy counts agree well with the fit to 
the data of Metcalfe
et al. (1995) and show a strong departure at $R = 23.5$ indicating
the completeness in our data. The bottom panel shows the percentage deviation 
of each field sight-line 
about the mean-B2547 value. The symbols are as defined in 
the top seven panels. The solid line represents the expected range of deviation
based on Poisson counting statistics.}
\end{figure}

\begin{figure}[p]
\vspace{-2.0cm}
\centerline{{\psfig{file=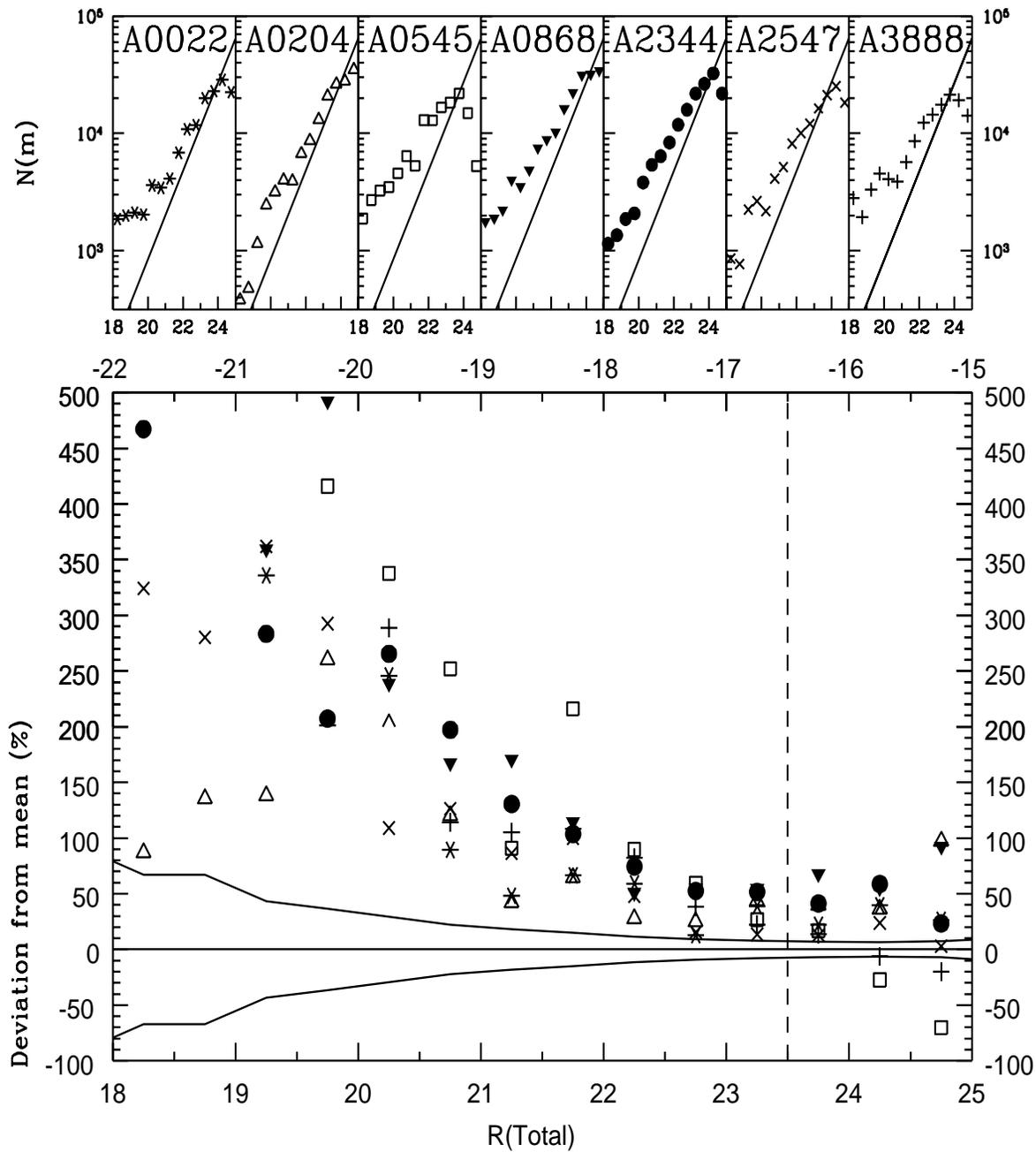,height=180mm,width=160mm}}}
\caption{As for Figure 3 the top panels show the individual
galaxy counts for each cluster sight-line. 
The lower panel shows the excess
number of galaxies over the mean field counts. The important point to 
notice is that at all magnitudes and for all clusters an
excess of galaxies is seen. The top axis shows the approximate absolute
magnitudes assuming a mean distance modulus for all clusters of 40.0.}
\end{figure}

\begin{figure}[p]
\centerline{{\psfig{file=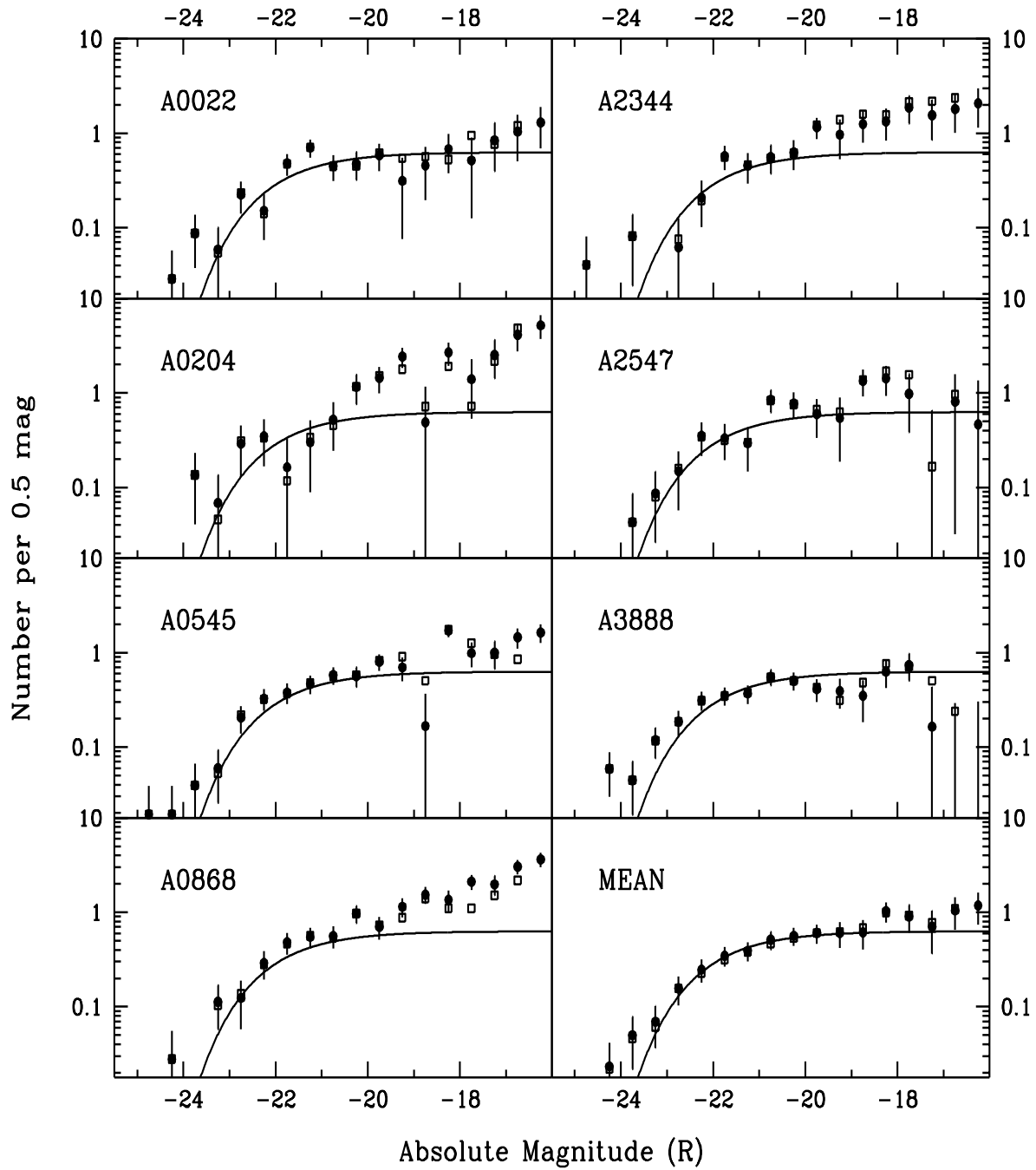,height=190mm,width=160mm}}}
\caption{The recovered luminosity distributions for the seven clusters
and their mean.
For the cluster A2547 no appropriate nearby field sight-line 
exists and the nearest field 
counts (B0022s) were used for the faint galaxy subtraction.
Solid dots indicate the recovered LDs with errors while the open squares
show the recovered LDs using an alternate background subtraction, see 
text}
\end{figure}

\begin{figure}[p]
\centerline{{\psfig{file=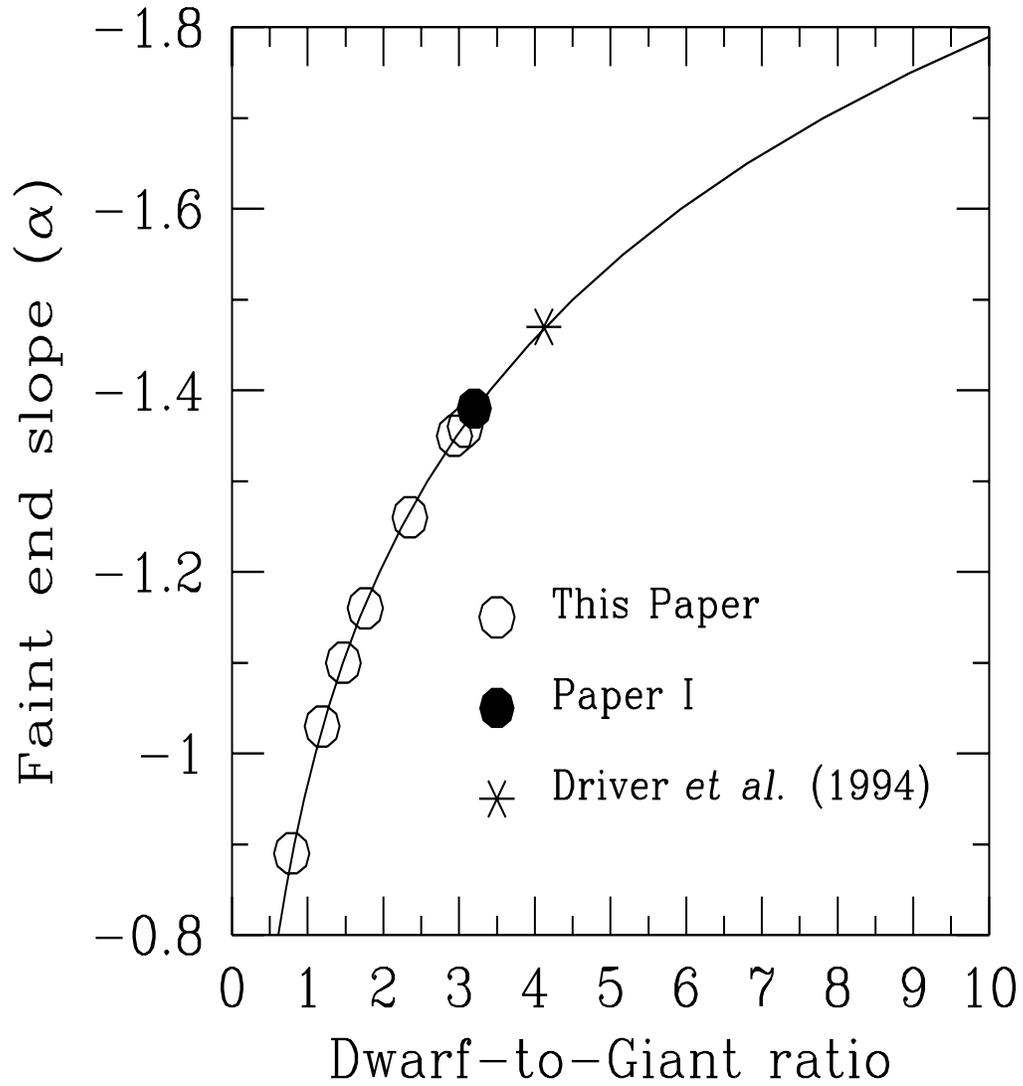,height=180mm,width=150mm}}}
\caption{The relationship between the dwarf-to-giant ratio and $\alpha$ 
the faint-end slope parameter adopting a standard Schechter function with 
$M_{R}^{*} =-22.5$.}

\end{figure}

\begin{figure}[p]
\centerline{{\psfig{file=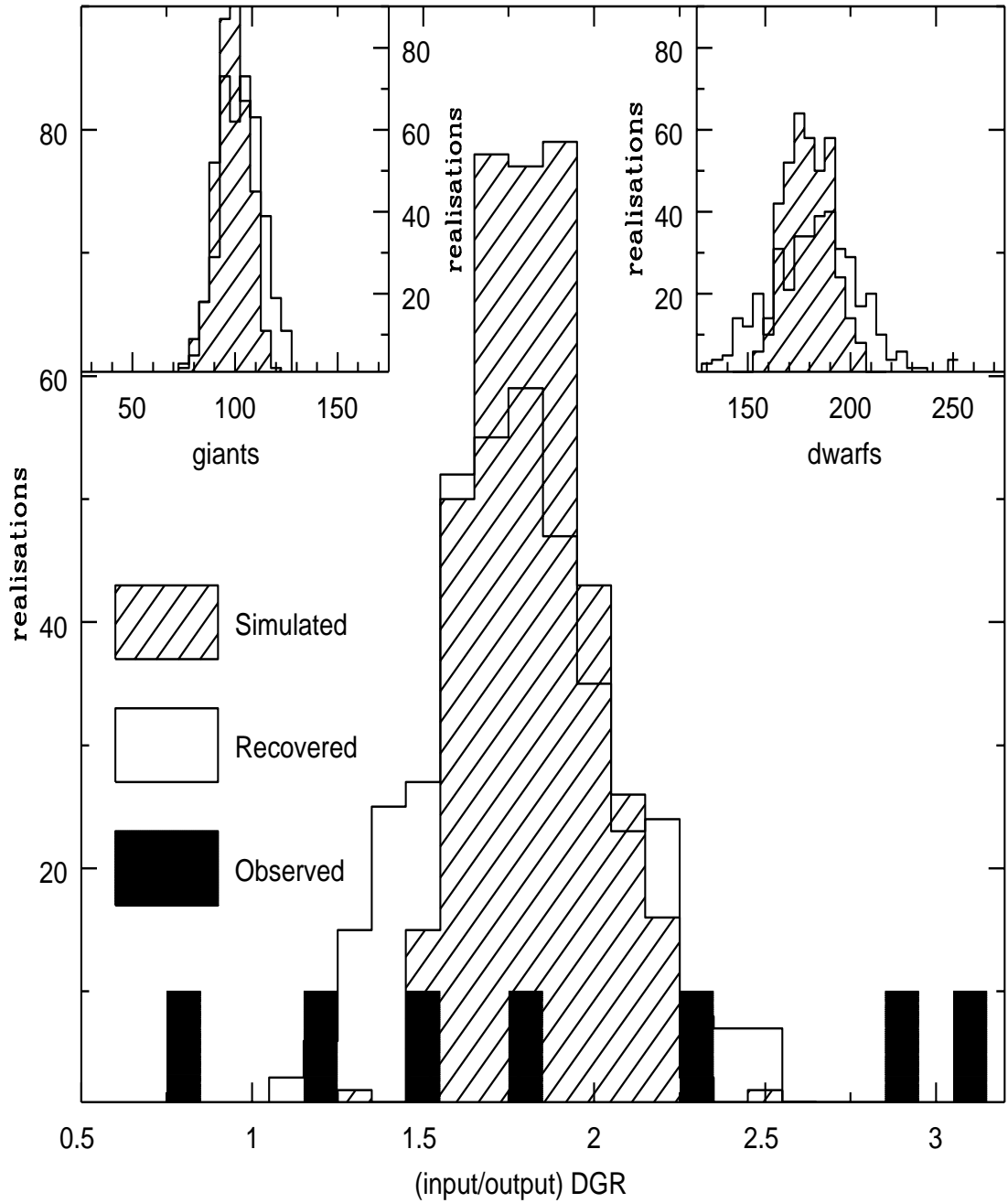,height=190mm,width=160mm}}}
\caption{Monte-Carlo simulations of 400 clusters to assess the errors in the 
measurement of a individual cluster's DGR}.
\end{figure}

\begin{figure}[p]
\centerline{{\psfig{file=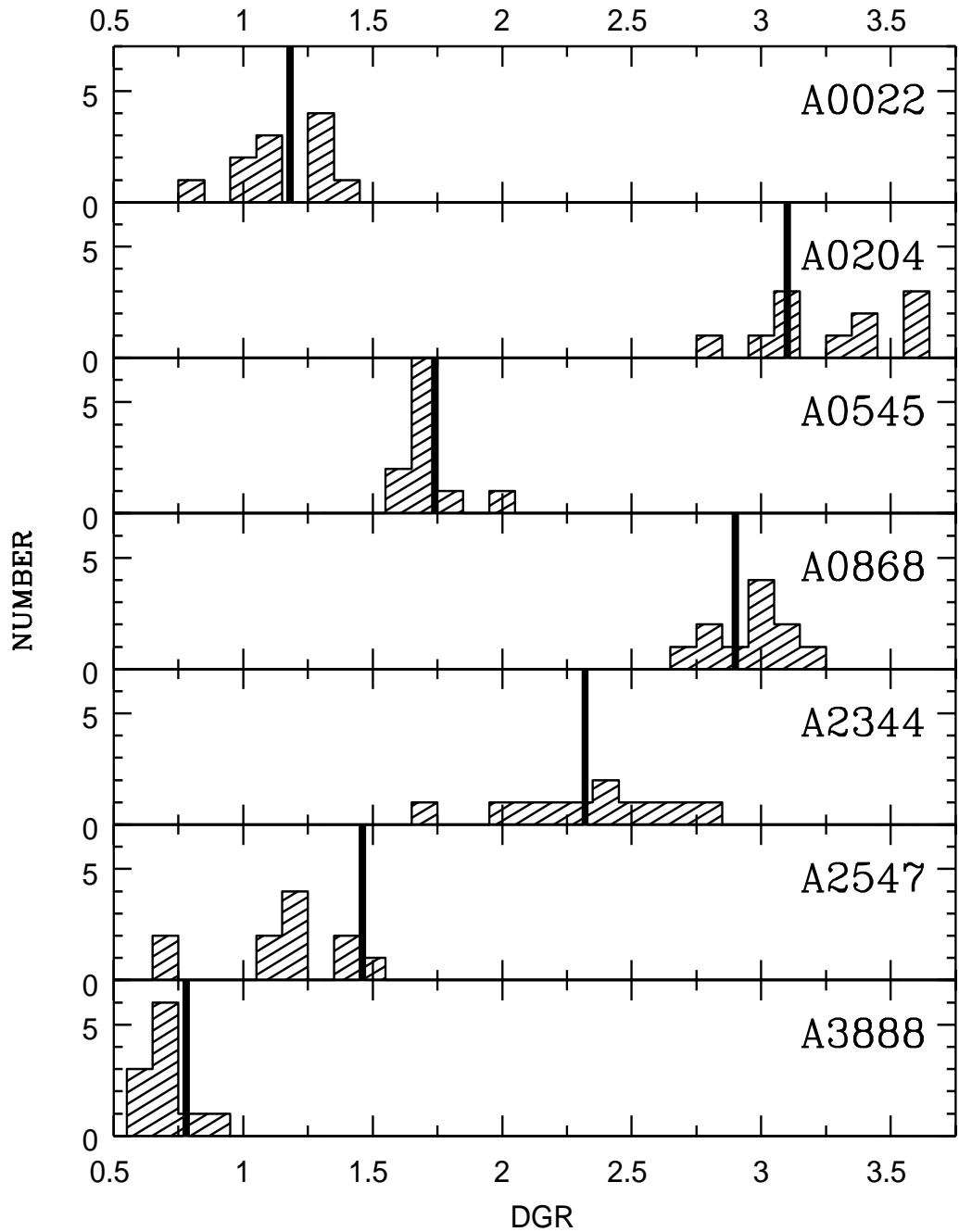,height=190mm,width=180mm}}}
\caption{Monte-Carlo simulations for each of the clusters to assess the 
reliability of the results. Each panel shows the resultant DGR distribution
after 11 simulations, if the derived characteristics for that cluster are used
as imput to our simulation software.}
\end{figure}

\begin{figure}[p]
\centerline{{\psfig{file=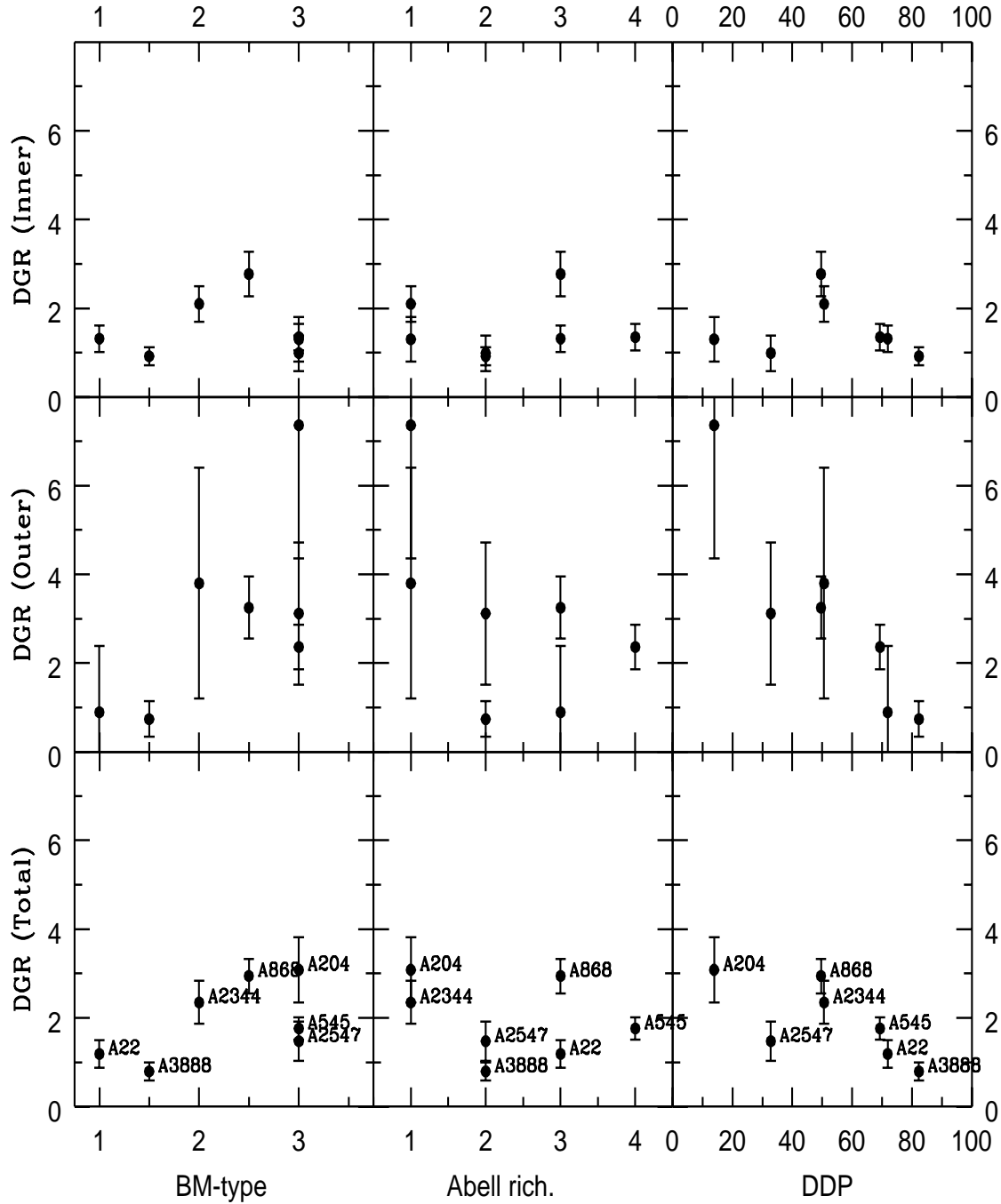,height=190mm,width=160mm}}}
\caption{The DGR statistic as a function of Bautz-Morgan type (left), Abell
richness (centre) and Dressler density parameter (right) for the innermost 1 
Mpc$^{2}$ (top), the remaining outer 
region outside of the central 1 Mpc$^{2}$ (middle), and
the entire field-of-view (bottom).}
\end{figure}

\begin{figure}[p]
\centerline{{\psfig{file=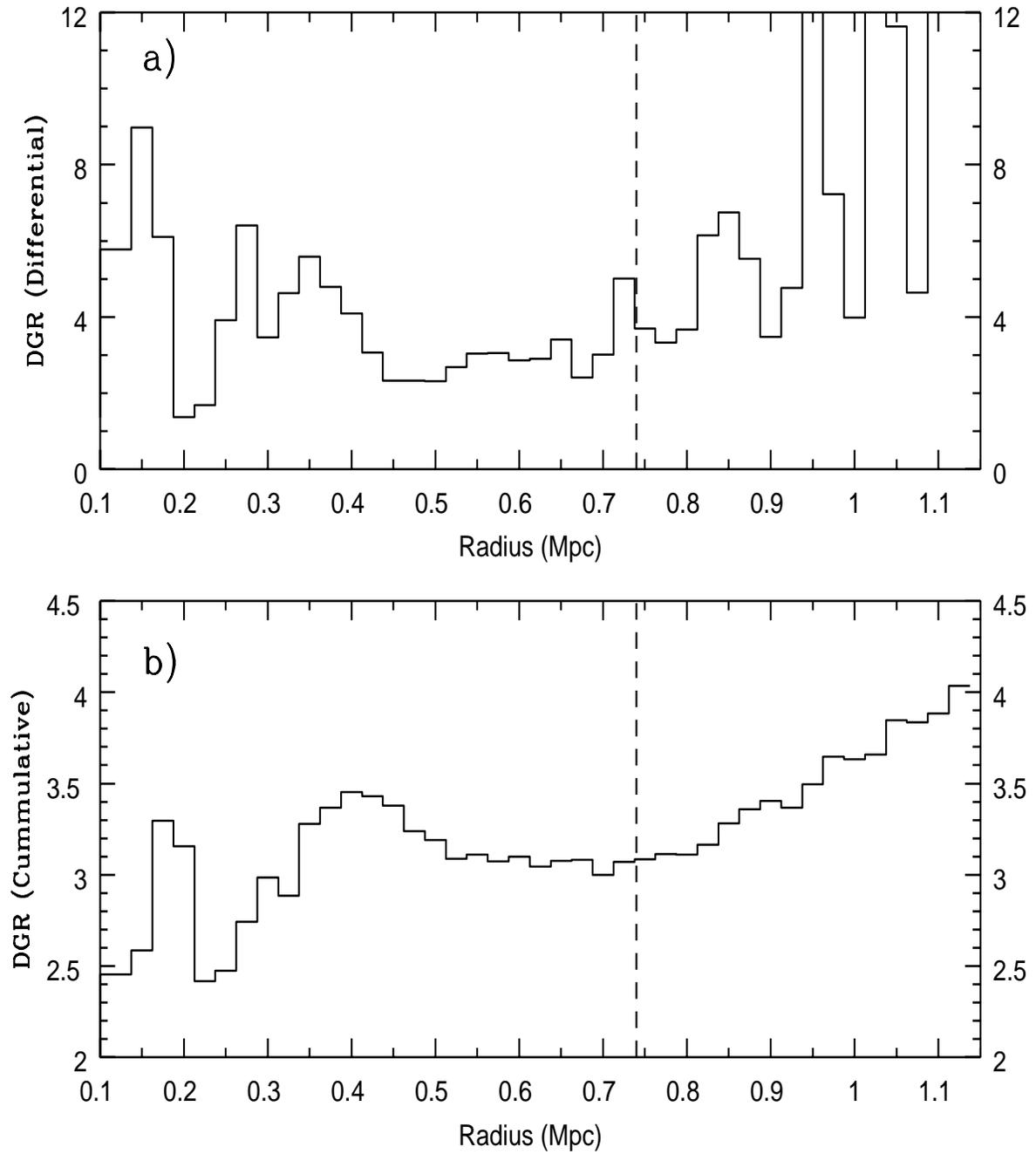,height=190mm,width=160mm}}}
\caption{The differential (a) and cumulative (b) DGR for A2554 as a function of
radius.}
\end{figure}

\begin{figure}[p]
\centerline{{\psfig{file=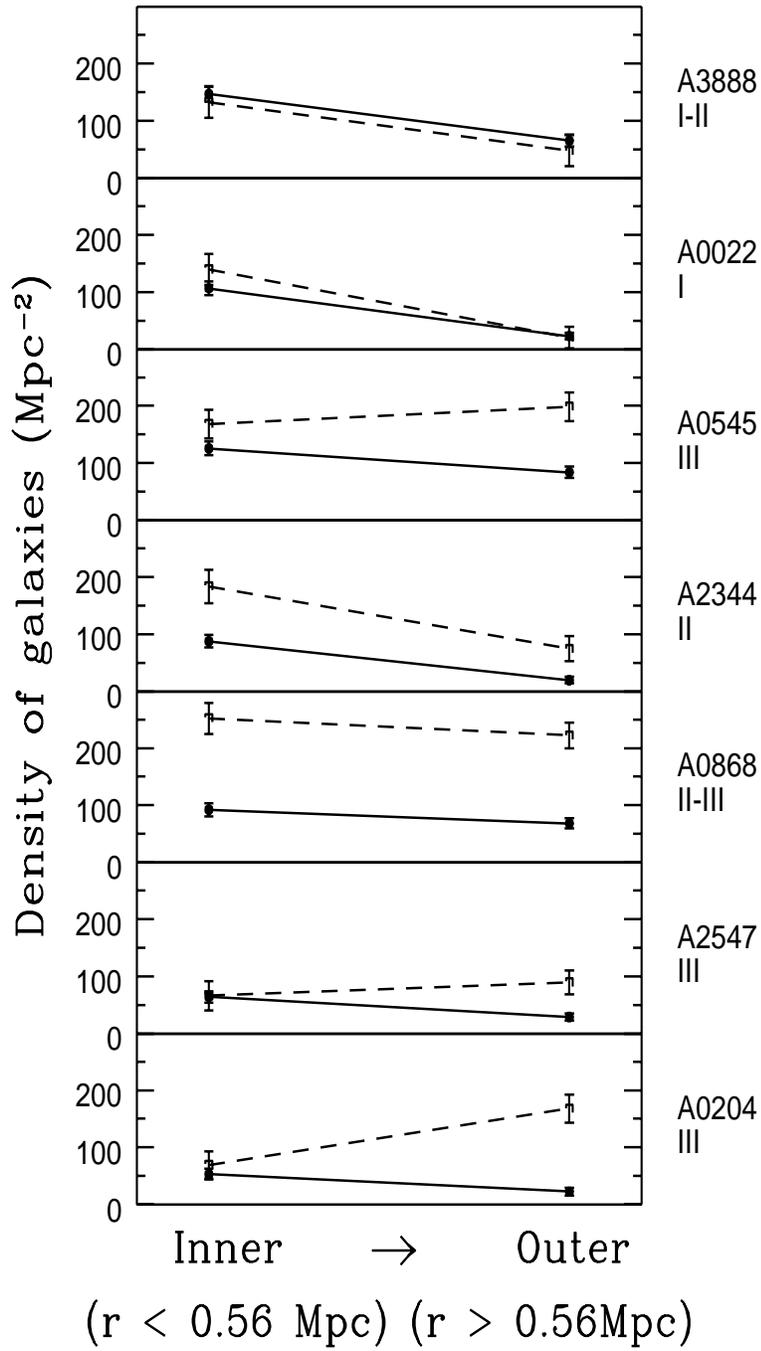,height=190mm,width=160mm}}}
\caption{The projected density of giants (solid) and dwarfs (dashed) for
the inner ($r < 0.56$ Mpc) and outer ($r > 0.56$ Mpc) regions of our seven 
clusters. The panels are ordered from top to bottom according to their
associated DDPs, Bautz-Morgan class is shown along the right-hand side.}
\end{figure}

\end{document}